\documentclass[10pt,journal,compsoc]{IEEEtran}
\ifCLASSINFOpdf
\else
\fi
\ifCLASSOPTIONcompsoc
  \usepackage[nocompress]{cite}
\else
  \usepackage{cite}
\fi
\usepackage[utf8]{inputenc} 
\usepackage[T1]{fontenc}    
\usepackage{hyperref}       
\usepackage{url}            
\usepackage{booktabs}       
\usepackage{amsfonts}       
\usepackage{nicefrac}       
\usepackage{microtype}      
\usepackage[bottom]{footmisc}
\usepackage{multicol}
\usepackage{algorithmicx}

\usepackage{graphicx} 
\usepackage{subcaption}
\usepackage{color}
\usepackage[dvipsnames]{xcolor}
\usepackage{listings}
\usepackage{soul}
\usepackage{algorithm}
\usepackage[noend]{algpseudocode}
\usepackage{tikz}
\usetikzlibrary{arrows,positioning,automata,calc, bayesnet}

\usepackage{times}

\usepackage{amssymb}
\usepackage{amsmath,bm}
\usepackage{xcolor}
\usepackage{listings}

\usepackage{mathtools}

\usepackage{wrapfig}
\usepackage{dblfloatfix}
\usepackage{hyperref}
\usepackage{float}
\usepackage{multirow}

\usepackage{acro}

\makeatletter
\renewcommand\footnotesize{%
   \@setfontsize\footnotesize\@ixpt{11}%
   \abovedisplayskip 8\p@ \@plus2\p@ \@minus4\p@
   \abovedisplayshortskip \z@ \@plus\p@
   \belowdisplayshortskip 4\p@ \@plus2\p@ \@minus2\p@
   \def\@listi{\leftmargin\leftmargini
               \topsep 4\p@ \@plus2\p@ \@minus2\p@
               \parsep 2\p@ \@plus\p@ \@minus\p@
               \itemsep \parsep}%
   \belowdisplayskip \abovedisplayskip
}
\makeatother

\algnewcommand\algorithmicon{\textbf{Each}}
\algnewcommand\algorithmicond  {\textbf{Process}}
\algnewcommand\algorithmicondd  {\textbf{All processes of }}

\algnewcommand\algorithmicfrom{\textbf{}}
\algnewcommand\algorithmicperform{\textbf{does:}}
\algnewcommand\algorithmicperformd{\textbf{does:}}
\algnewcommand\algorithmicperformdd{\textbf{collectively do:}}
\algnewcommand{\LineComment}[1]{\State \(\blacktriangleright\) \textit{#1}}
\algnewcommand\algorithmicforeach{\textbf{for each}}
\algdef{S}[FOR]{ForEach}[1]{\algorithmicforeach\ #1\ \algorithmicdo}

\algblockdefx[ON]{On}{EndOn}[2]
  {\algorithmicon\ #1\ \algorithmicfrom\ #2\ \algorithmicperform}
  {\algorithmicend\ \algorithmicon}
\algblockdefx[OND]{Ond}{EndOnd}[1]
  {\algorithmicond\ #1 \algorithmicperformd}
  {\algorithmicend\ \algorithmicon}
\algblockdefx[ONDD]{Ondd}{EndOndd}[1]
  {\algorithmicondd\ #1 \algorithmicperformdd}
  {\algorithmicend\ \algorithmicon}

\makeatletter
\ifthenelse{\equal{\ALG@noend}{t}}%
  {\algtext*{EndOn}}
  {}%
\ifthenelse{\equal{\ALG@noend}{t}}%
  {\algtext*{EndOndd}}
  {}%
\makeatother

\makeatletter
\ifthenelse{\equal{\ALG@noend}{t}}%
  {\algtext*{EndOnd}}
  {}%
\makeatother

\def\node#1{{\emph{\textsf{#1}}}}

\providecommand{\keywords}[1]
{
  \small	
  \textbf{\textit{Keywords---}} #1
}

\def\eqref#1{Eq.~(\ref{#1})}
\def\figref#1{Fig.~(\ref{#1})}

\def\algref#1{Alg.~\ref{#1}}

\DeclareFontFamily{U}{FdSymbolC}{}
\DeclareFontShape{U}{FdSymbolC}{m}{n}{<-> s * FdSymbolC-Book}{}
\DeclareSymbolFont{fdarrows}{U}{FdSymbolC}{m}{n}
\DeclareMathSymbol{\leftinhibits}{\mathrel}{fdarrows}{"AC}
\DeclareMathSymbol{\rightinhibits}{\mathrel}{fdarrows}{"AD}
\DeclareMathSymbol{\longleftinhibits}{\mathrel}{fdarrows}{"C6}
\DeclareMathSymbol{\longrightinhibits}{\mathrel}{fdarrows}{"C7}

\DeclareAcronym{ett}{
    short = ETT,
    alt   = \text{ETT},
    long  = effect of the treatment on the treated.
}

\DeclareAcronym{ite}{
    short = ITE,
    alt   = \text{ITE},
    long  = individual treatment effect
}

\DeclareAcronym{ate}{
    short = ATE,
    alt   = \text{ATE},
    long  = average treatment effect
}

\DeclareAcronym{raf}{
    short = RAF,
    alt   = \text{RAF},
    long  = Rapidly Accelerated Fibrosarcoma
}
\DeclareAcronym{mek}{
    short = MEK,
    alt   = \text{MEK},
    long  = MAPK ERK kinase,
}
\DeclareAcronym{erk}{
    short = ERK,
    alt   = \text{ERK},
    long  = Extracellular signal-regulated kinases,
}
\DeclareAcronym{toci}{
    short = Toci,
    alt   = \text{Toci},
    long  = Tocilizumab,
}

\DeclareAcronym{prr}{
    short = PRR,
    alt   = \text{PRR},
    long  = Pathogen Recognition Receptors,
}

\DeclareAcronym{ace2}{
    short = ACE2,
    alt   = \text{ACE2},
    long  = Angiotensin Converting Enzyme 2,
}

\DeclareAcronym{agtr1}{
    short = AGTR1,
    alt   = \text{AGTR1},
    long  = Angiotensin II Receptor 1,
}

\DeclareAcronym{adam17}{
    short = ADAM17,
    alt   = \text{ADAM17},
    long  = A Disintegrin And Metalloproteinase 17,
}
\DeclareAcronym{tnf}{
    short = TNF$\alpha$,
    alt   = \text{TNF$\alpha$},
    long  = Tumor Necrosis Factor $\alpha$,
}
\DeclareAcronym{il6stat3}{
    short = IL6-STAT3,
    alt   = \text{IL6-STAT3},
    long  = Interleukin 6 STAT3 Complex,
}

\DeclareAcronym{egf}{
    short = EGF,
    alt   = \text{EGF},
    long  = Epidermal Growth Factor,
}
\DeclareAcronym{ards}{
    short = cytokine storm,
    alt   = \text{CytokineStorm},
    long  = Cytokine Release Syndrome
}
\DeclareAcronym{nfkb}{
    short = NF-$\kappa$B,
    alt   = \text{NF-$\kappa$B},
    long  = nuclear factor kappa-light-chain-enhancer of activated B cell,
}
\DeclareAcronym{il6r}{
    short = mIL6R,
    long  = membrane-bound Interleukin 6 receptor,
}
\DeclareAcronym{sil6r}{
    short = sIL6R$\alpha$,
    alt   = \text{sIL6R$\alpha$},
    long  = soluble Interleukin 6 receptor,
}
\DeclareAcronym{il6amp}{
    short = IL6-AMP,
    alt   = \text{IL6-AMP},
    long  = Interleukin 6 Amplifier,
}
\DeclareAcronym{il6}{
    short = IL6,
    alt   = \text{IL6},
    long  = Interleukin 6,
}
\DeclareAcronym{il6st}{
    short = IL6ST,
    alt   = \text{IL6ST},
    long  = Interleukin 6 Signal Transducer
}

\DeclareAcronym{stat3}{
    short = STAT3,
    alt   = \text{STAT3},
    long  = Signal Transducer and Activator of Transcription 3,
}

\DeclareAcronym{egfr}{
    short = EGFR,
    alt   = \text{EGFR},
    long  = Epidermal Growth Factor Receptor,
}

\DeclareAcronym{sars2}{
    short = SARS-CoV-2,
    alt   = \text{SARS-CoV-2},
    long  = severe acute respiratory syndrome coronavirus 2,
}

\DeclareAcronym{bionlp}{
    short = BioNLP,
    long  = natural language processing of biomedical literature
}

\DeclareAcronym{ang}{
    short = Angiotensin II,
    alt   = \text{Angiotensin II},
    long  = Angiotensin II,
}

\DeclareAcronym{AF}{
    short = AF,
    long  = Activity Flow,
    long-plural-form = Activity Flows,
    cite  = le2009systems,
}
\DeclareAcronym{sbml}{
    short = SBML,
    long  = Systems Biology Markup Language,
    cite  = hucka2018systems,
}

\DeclareAcronym{qual}{
    short = SBML-qual,
    long  = \ac{sbml} Qualitative Models Package,
    cite  = chaouiya_2013,
}

\DeclareAcronym{biopax}{
    short = BioPAX,
    long  = Biological Pathway Exchange Language,
    cite  = demir_2010,
}
\DeclareAcronym{BEL}{
    short = BEL,
    long  = Biological Expression Language,
    cite  = Slater2014,
}
\DeclareAcronym{covid19}{
    short = COVID-19,
    long  = coronavirus disease 2019
}
\DeclareAcronym{igf}{
    short = IGF,
    long  = insulin-like growth factor
}
\DeclareAcronym{INDRA}{
    short = INDRA,
    long  = Integrated Dynamical Reasoner and Assembler,
    cite  = gyori2017word,
}
\DeclareAcronym{ODE}{
    short = ODE,
    short-plural-form = ODEs,
    long  = ordinary differential equation,
    long-plural-form = ordinary differential equations,
}
\DeclareAcronym{PDE}{
    short = PDE,
    long  = partial differential equation,
    long-plural-form = partial differential equations,
}
\DeclareAcronym{NFKB}{
    short = NF-$\kappa$B,
    long  = nuclear factor kappa-light-chain-enhancer of activated B cells,
}
\DeclareAcronym{NPA}{
    short = NPA,
    long  = Network Perturbation Amplitude,
    cite  = Martin2014
}
\DeclareAcronym{MAPK}{
    short = MAPK,
    long  = mitogen-activated protein kinase,
}
\DeclareAcronym{mRNA}{
    short = mRNA,
    long  = messenger ribonucleic acid,
    long-plural-form = messenger ribonucleic acids,
}
\DeclareAcronym{PD}{
    short = PD,
    long  = Process Description,
    long-plural-form = Process Descriptions,
    cite  = le2009systems,
}
\DeclareAcronym{RCR}{
    short = RCR,
    long  = Reverse Causal Reasoning,
    cite  = Catlett2013
}
\DeclareAcronym{sarscov2}{
    short = SARS-CoV-2,
    long  = severe acute respiratory syndrome coronavirus 2
}
\DeclareAcronym{SBGN}{
    short = SBGN,
    long  = Systems Biology Graphical Notation,
    cite  = le2009systems,
}

\DeclareAcronym{SCM}{
    short            = SCM,
    long             = structural causal model,
    long-plural-form = structural causal models,
}
\DeclareAcronym{SDE}{
    short = SDE,
    long  = stochastic differential equation,
    long-plural-form = stochastic differential equations,
}
\DeclareAcronym{bel2scm}{
    short = \textsc{bel2scm},
    long  = Biological Expression Language to Structural Causal Models,
    alt   = https://github.com/bel2scm
}     
\DeclareAcronym{query2bel}{
    short = \textsc{query2bel},
    long  = Causal query to Biological Expression Language,
    alt   = https://github.com/bel2scm
}     

\DeclareAcronym{cord19}{
    short = CORD-19,
    long  = COVID-19 Open Research Dataset,
    cite  = luwang_2020,
}     

\DeclareAcronym{biodati}{
    short = BioDati,
    long  = BioDati Inc.,
}
\DeclareAcronym{covid19kg}{
    short = Covid19kg,
    long  = Fraunhofer Covid-19 Knowledge Graph,
}
\DeclareAcronym{sigscm}{
    short = $\aca{SCM}^\sigma$,
    long  = sigmoidal \acl{SCM}
}
\DeclareAcronym{mmscm}{
    short = $\aca{SCM}^{MM}$,
    long  = \acl{SCM} with a Michaelis-Menten functional form
}

\begin{document}
%
\title{Leveraging Structured Biological Knowledge for Counterfactual Inference: a Case Study of Viral Pathogenesis}



\author{Jeremy~Zucker$^{1*}$, 
        Kaushal~Paneri$^{2*}$, 
        Sara~Mohammad-Taheri$^{3*}$,\\
        Somya Bhargava$^{3}$, 
        Pallavi Kolambkar$^{3}$, 
        Craig Bakker$^1$, 
        Jeremy Teuton$^1$, 
        Charles Tapley Hoyt$^4$, 
        Kristie Oxford$^1$, 
        Robert Ness$^5$
        and~Olga~Vitek$^{3\dag}$}

\thanks{ 
$^*$Equal contribution\\
$^1$ Pacific Northwest National Laboratory, Richland, WA\\
$^2$ Microsoft, Redmond, WA\\
$^3$  Northeastern University, Boston, MA\\
$^4$ Enveda Biosciences, Bonn, Germany\\
$^5$ Altdeep, Boston, MA\\
$^{\dag}$ Corresponding author: ovitek@neu.edu
}

\maketitle

\begin{abstract}
Counterfactual inference is a useful tool for comparing outcomes of interventions on complex systems.
It requires us to represent the system in form of a structural causal model, complete with a causal diagram, probabilistic assumptions on exogenous variables, and functional assignments.
Specifying such models can be extremely difficult in practice.
The process requires substantial domain expertise, and does not scale easily to large systems, multiple systems, or novel system modifications.
At the same time, many application domains, such as molecular biology, are rich in structured causal knowledge that is qualitative in nature.
This manuscript proposes a general approach for querying a causal biological knowledge graph, and converting the qualitative result into a quantitative structural causal model that can learn from data to answer the question.
We demonstrate the feasibility, accuracy and versatility of this approach using two case studies in systems biology. 
The first demonstrates the appropriateness of the underlying assumptions and the accuracy of the results.
The second demonstrates the versatility of the approach by querying a knowledge base for the molecular determinants of a \ac{sars2}-induced cytokine storm, and performing counterfactual inference to estimate the causal effect of medical countermeasures for severely ill patients.
\end{abstract}

%
\IEEEpeerreviewmaketitle

\section{Introduction}\label{sec:introduction}





Each time a cell senses changes in its environment, it marshals a
complex choreography of molecular interactions to initiate an
appropriate response.
When a virus infects the cell, this delicate balance is disrupted and
can result in a cascade of systemic failures leading to disease.
In particular, \acf{sarscov2}, the novel pathogen responsible for the COVID-19 pandemic, has a
complex etiology that differs in subtle and substantial ways from
previously studied viruses.
To make informed decisions about the risk that a new pathogen presents, it is imperative to rapidly predict the determinants of pathogenesis and identify potential targets for medical countermeasures.
Current solutions for this task include systems biology data-driven models, which correlate biomolecular expression to pathogenicity, but cannot go beyond associations in the data to reason about causes of the disease~\cite{pezeshki_2019,pedragosa_2019}. 
Alternatively, hypothesis-driven mathematical models capture causal relations, but are hampered by limited parameter identifiability and predictive power~\cite{nguyen_2016,arazi_2013}.

We argue that counterfactual inference~\cite{pearl2009causal} helps bridge the gap between data-driven and hypothesis-driven approaches. 
It enables questions of the form: ``Had we known the eventual outcome of a patient, what would we have done differently?''
At the heart of counterfactual inference is a formalism known as a \acf{SCM}~\cite{peters2017elements,pearl2009causal}.
It represents prior domain knowledge in terms of causal diagrams, assumes a probability distribution on exogenous variables, and assigns a deterministic function to endogenous variables.
SCM are particularly attractive in systems biology, where structured domain knowledge is extracted from the biomedical literature and is readily available through advances in natural language processing~\cite{Allen2008,McDonald2000,Valenzuela-Escarcega2018}, large-scale automated assembly systems~\cite{gyori2017word}, and semi-automated curation workflows~\cite{hoyt2019}.
This knowledge is curated by multiple organizations~\cite{Cerami2011,Fabregat2018,Kanehisa2017,Perfetto2016,Slenter2018} and stored in structured knowledge bases~\cite{demir_2010,hucka2018systems,le2009systems,Slater2014}.  
It can be brought to bear for answering causal questions regarding SARS-CoV-2. 

This manuscript contributes a three-part algorithm that leverages existing structured biological knowledge to answer counterfactual questions about viral pathogenesis.
Algorithm 1 formalizes biologically relevant questions as queries to an existing causal knowledge graph.
Algorithm 2 converts the query result into a structural causal model.
Algorithm 3  operationalizes the counterfactual inference by interrogating the model with the observed data to estimate a causal effect.

We illustrate the benefits of this approach using two case studies.
Case study 1 illustrates the increased precision of counterfactual estimates, as compared to the ODE- and SDE-based forward simulation, in a situation with known ground truth mechanisms of data generation.
Case study 2 demonstrates the automated construction of an SCM and the value of counterfactual reasoning in novel situations with limited treatment options (as is the case for SARS-CoV-2).
It shows that counterfactual inference enables more precise predictions regarding who would be likely to survive without receiving treatment, who would be likely to die even if they did receive treatment, and who would likely survive only if they received treatment.

\section{Background}\label{sec:background}
\noindent {\textbf{Biological signaling pathways}}
Signaling pathways are composed of entities that engage in activities~\cite{machamer_2000}.
Examples of entities are proteins and metabolites, but also higher level biological processes such as an immune response.
Activities are the producers of change.
Examples include catalytic activity, kinase activity, or transcriptional activity.

The basic unit of causality in signaling pathways is a directed molecular interaction, where the activity of an upstream molecule increases or decreases the activity of a downstream molecule. 
For example, the \acf{MAPK} intracellular signaling pathway is a causal chain of directed molecular interactions shown  in~\eqref{mapkNet} 
\begin{eqnarray}
    \label{mapkNet}
    a(S_1) \rightarrow kin(p(Raf)) \rightarrow kin(p(Mek)) \rightarrow kin(p(Erk))
\end{eqnarray}
The interactions transmit information about a stimulus at the cell surface to the nucleus, where proteins called transcription factors activate an appropriate biological process~\cite{Li2017}.  
A causal diagram of \ac{MAPK} consists of a signaling molecule $S_1$ and three proteins $Raf$, $Mek$, and $Erk$, each of which engage in kinase activity.
We represent signaling molecule abundance with $a()$, protein abundance with $p()$ and the kinase activity of a protein with $kin()$.
In the case of MAPK, the abundance or activity of an upstream entity causes the abundance or activity of a downstream entity to increase, and is represented with a sharp edge $\rightarrow$.
The diagram is a abstraction showing that the abundance of the signaling molecule  $S_1$ increases the kinase activity of $Raf$, which increases the kinase activity of $Mek$, which increases the kinase activity of $Erk$. 
In other cases, if the abundance or activity of an upstream entity causes the abundance or activity of a downstream entity to decrease, we represent this with a blunt edge $\rightinhibits$.

\medskip \noindent {\textbf{Viral dysregulation}}
Viral disruptions of a signaling pathway take form of overactivation or repression of its activities.
For example, by amplifying the release of intercellular signaling molecules that overstimulate the immune response, known as \acf{ards}, a virus can cause severe system-level cellular damage.

\medskip \noindent {\textbf{Quantitative modeling of biological processes with ODE/SDE}}
Temporal dynamics of biological processes can be expressed quantitatively using ordinary (or stochastic) differential equations.
A small number of high quality, validated models have been published in the literature and stored in a computable form in repositories such as Biomodels~\cite{chen2010modeling,gratie2013ode}.
For example, the \ac{MAPK} signaling pathway in~\eqref{mapkNet} is well characterized. We denote $R(t)$, $M(t)$, and $E(t)$ as the respective amounts of active $Raf$, $Mek$, and $Erk$ at time $t$;
We denote $T_{R}$, $T_{M}$, and $T_{E}$ as their total amounts, which we assume do not change during the considered timeframe;
$v^\text{act}_R$, $v^\text{inh}_R$, $v^\text{act}_M$, $v^\text{inh}_M$, $v^\text{act}_E$, and $v^\text{inh}_E$ are experimentally derived activation or inhibition kinetic rate constants; and $S_1$ is the amount of the input signal.
The system of \acp{ODE} is specified as follows \cite{bianconi2012computational, kim2010pathological}:

\begin{small}
\label{mapkODE}
\begin{align}
        \label{mapkODE}
        \frac{\mathrm{d}R}{\mathrm{d}t} &= v^{\text{act}}_{R} S_1 (T_{R}-R(t)) - v^{\text{inh}}_{R}R(t)  \\
        \frac{\mathrm{d}M}{\mathrm{d}t} &=  \frac{(v^{\text{act}}_{M})^2}{v^{\text{inh}}_{M}}R(t)^2(T_{M} - M(t)) - v^{\text{act}}_{M}R(t)M(t) - v^{\text{inh}}_{M}M(t) \nonumber \\
        \frac{\mathrm{d}E}{\mathrm{d}t} &=  \frac{(v^{\text{act}}_{E})^2}{v^{\text{inh}}_{E}}M(t)^2(T_{E} - E(t)) - v^{\text{act}}_{E}M(t)E(t) - v^{\text{inh}}_{E}E(t) \nonumber
\end{align}
\end{small}

Given initial conditions, forward simulations from the \acp{ODE} can be used to generate the temporal trajectories of the amounts of activated proteins , such as $R(t)$, $M(t)$, and $E(t)$ in the \ac{MAPK} example.
In this manuscript we refer to such simulated data as \textit{observational data}.
We define an \textit{ideal intervention} as an event that fixes the amount of an activated protein.
For example, if we fix the kinase acivity of $Mek$ at $M(t) = m$, the second equality $\frac{\mathrm{d} M}{\mathrm{d} t}$ in~\eqref{mapkODE} becomes zero.
We can simulate data from \eqref{mapkODE} with $\frac{\mathrm{d} M}{\mathrm{d} t}=0$, and refer to these as \textit{interventional data}.
Contrasting observational and interventional data helps evaluate the outcome of the intervention~\cite{ness2019integrating}.

The deterministic \ac{ODE} ignore the fact that at low concentration, stochasticity becomes a significant factor in determining the reaction \cite{paneri2019integrating}.
As the collisions between molecules participating in biochemical process become stochastic, a stochastic model is required.
In contrast to \ac{ODE}, a stochastic differential equation model or \ac{SDE} specifies biological process as a random process.
For example, in the case of \ac{MAPK}, the random process of the reaction $Mek \rightarrow Erk$ is specified with
\begin{eqnarray}
\frac{dP_{E} (t)}{dt} = g_{E}(t, v^{act}_E, v^{inh}_E, M(t)),\  E(0) = e_0
\end{eqnarray}
where $P_E(t)$ is marginal probability density of $E(t)$, function $g_{E}$ determines the probability of a state change between $E(t)$ and $E(s), s > t$, $e_0$ is initial condition, and $M(t)$ is the value of its parent Mek at $t$.
Once \acl{SDE} are fully specified, one can use, e.g. Gillespie's stochastic simulation algorithm ~\cite{gillespie1977exact} to simulate observational and interventional data, and evaluate the outcomes of interventions.

Unfortunately, even simple \acp{ODE} such as the one in the \ac{MAPK} example are difficult to build \textit{de novo}. This is nearly impossible for novel and poorly studied systems that lack the existence or findability of experimental information describing the structure or boundaries of the process, kinetic equations governing their dynamics~\cite{jha2012exploring}, rate constants for these equations, or rules governing each agents' states and functions.

\medskip \noindent {\textbf{Equilibrium enzyme kinetics}} Simpler and more general quantitative models can be specified when a reaction reaches the state of chemical equilibrium~\cite{alon2019introduction}. One commonly used such model is \textit{Hill function} in the form of 
\begin{eqnarray}
X = \beta \frac{\mathbf{PA}_X^n}{K^n + \mathbf{PA}_X^n}
\end{eqnarray}
where $X$ is the abundance of a protein in a causal diagram (such as $Erk$ in~\eqref{mapkNet}), $\mathbf{PA}_X$ is the set of its parents, $n$ is a parameter interpreted as the number of ligand binding sites of the protein, and $\beta$ is the total number of molecules of the protein. A special and frequently used case of the Hill function, called \textit{Michaelis-Menten} function, occurs when $n=1$. Although simple to use, these models are deterministic, and do not describe the stochasticity that is a distinctive property of biological systems at low concentrations.

\begin{figure}[t!]
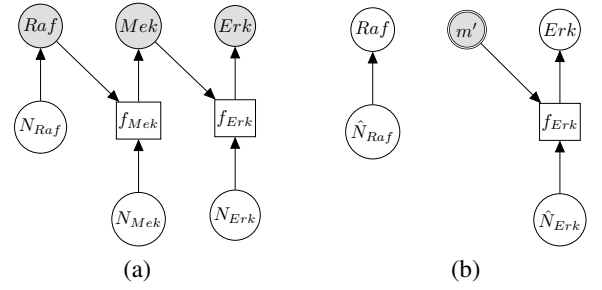

\centering
\begin{tabular}{ccc}
\scalebox{.7}{\tikz{
     \node[obs] (Raf) {$Raf$};%
     \node[obs,right=of Raf] (Mek) {$Mek$}; %
     \node[obs,right=of Mek] (Erk) {$Erk$}; %
     \node[latent,,rectangle,below=of Erk] (fErk) {$f_{Erk}$}; %
    \node[latent,,rectangle,below=of Mek] (fMek) {$f_{Mek}$}; %
    \node[latent, below =of Raf] (NRaf) {$N_{Raf}$} ; %
    \node[latent, below =of fErk] (NErk) {$N_{Erk}$} ; %
    \node[latent, below =of fMek] (NMek) {$N_{Mek}$} ; %
    \edge{Raf}{fMek};
     \edge {fMek}{Mek};
     \edge {Mek} {fErk};
     \edge{fErk} {Erk};
     \edge {NRaf}{Raf};
     \edge {NErk} {fErk};
     \edge {NMek} {fMek}
     }}& $~~~$ &
     \scalebox{.7}{\tikz{
     \node[latent] (Raf) {$Raf$};%
     \node[obs,accepting,right=of Raf] (Mek) {$m'$}; %
     \node[latent,right=of Mek] (Erk) {$Erk$}; %
     \node[latent,,rectangle,below=of Erk] (fErk) {$f_{Erk}$}; %
    \node[latent, below =of Raf] (NRaf) {$\hat{N}_{Raf}$} ; %
    \node[latent, below =of fErk] (NErk) {$\hat{N}_{Erk}$} ; %
   
     \edge {Mek} {fErk};
     \edge{fErk} {Erk};
     \edge {NRaf}{Raf};
     \edge {NErk} {fErk};
     }}\\
     (a) & & (b)
\end{tabular}

\caption{\small \textbf{Causal modeling of MAPK signaling pathway} Circles are variables, double circles are variables intervened upon, squares are deterministic functional assignments, gray nodes are observed variables, and white nodes are hidden variables. (a) Structural causal model. $N_{Raf}$, $N_{Mek}$ and $N_{Erk}$ are statistically independent noise variables. Root node $Raf$ is only dependent on noise variable $N_{Raf}$. Non-root nodes $Mek$ and $Erk$ are dependent on their parent and on the associated noise variable. (b) Counterfactual model. The intervention fixes the count of phosphorylated $Mek$ to $m'$, such that $Mek$ is no longer dependent on $Raf$ and $N_{Mek}$. Given an observed data point, counterfactual inference infers the noise variables $\hat{N}_{Raf}$, and $\hat{N}_{Erk}$. \label{fig:mapkNet}}

\end{figure}

\medskip \noindent {\textbf{Modeling biological processes with structural causal models}}
The stochastic nature of biological processes at steady-state can be represented by an \ac{SCM} such as in ~\figref{fig:mapkNet} (a)~\cite{BongersMooij_1803.08784,ness2019integrating}.
\acp{SCM} represent the dependencies between a child node $X$ and its parents $\mathbf{PA}_X$ in terms of a deterministic function $X = f_X(\mathbf{PA}_X, N_X)$ called \textit{structural assignment}, and a noise variable $N_X$.
In~\figref{fig:mapkNet} (a), $f_{Mek}$ and $f_{Erk}$ are linear or non-linear structural assignments, and $N_{Raf}$, $N_{Mek}$, and $N_{Erk}$ are statistically independent noise variables with defined probability distributions
\begin{eqnarray}
Raf = N_{Raf};\ Mek = f_{Mek}(Raf ; N_{Mek})\\
Erk = f_{Erk}(Mek,N_{Erk}) \nonumber
\label{SCMBackground}
\end{eqnarray}
An ideal intervention in an \ac{SCM} is performed on a functional assignment. For example, an ideal intervention on $Mek$ sets $Mek = m^{\prime}$, defining a new \ac{SCM}
\begin{eqnarray}
    \label{IntervenedSCMBackground}
    & Raf = N_{Raf};  Mek = m^{\prime};  Erk = f_{Erk}(Mek,N_{Erk}) &
\end{eqnarray}
An ideal intervention can also be thought of as a process of mutilating the causal graph.
For example, intervening on $Mek$ eliminates its dependence upon $Raf$, and therefore the edge from $Raf$ to $Mek$ is removed as shown in \figref{fig:mapkNet}(b). 

\medskip \noindent {\textbf{Counterfactual inference with \ac{SCM}}
Beyond direct model-based predictions, \Acp{SCM} enable {\it counterfactual inference}, i.e., the process of inferring the unseen outcomes of a hypothetical intervention given data observed in absence of the intervention~\cite{pearl2009causal}.
In the context of \ac{SCM}, counterfactuals are defined as operations
\begin{eqnarray}
& Y_{do(T=t^{\prime})}(u) \triangleq Y_{M_{do(T=t^{\prime})}}(u) &
\end{eqnarray}
In other words, the outcome $Y$ that individual $u$ would have had she received treatment $t^{\prime}$ is defined as the value that $Y$ would have in a structural causal model $M$ mutilated to replace $T=f_T(\cdot)$ with $T=t^{\prime}$.

For example, in the MAPK signaling pathway, we may be interested in the counterfactual question: \textit{Having observed the kinase activities of  $Raf = r$, $Mek = m$, $Erk = e$, what would be the kinase activity of $Erk$ in a hypothetical experiment where the kinase activity of $Mek$ was fixed to $m^{\prime}$?}
This counterfactual query is more formally translated into
\begin{eqnarray}
    P (Erk_{do( Mek= m^{\prime})} | Raf = r, Mek = m, Erk = e) \label{eq:exampleCounterfactual}
\end{eqnarray}
The probability distribution in~\eqref{eq:exampleCounterfactual} is estimated with the following steps:
\begin{enumerate}
\item \textbf{Abduction:} Given observational data, estimate the posterior distribution of the noise variables.
In the MAPK example, we estimate the posterior distribution of the noise variables:
\begin{align*}\hat{N}_{Raf} = & \{N_{Raf} | Raf = r, Mek = m, Erk = r\} \\
\hat{N}_{Erk} = & \{N_{Erk} | Raf = r, Mek = m, Erk = r\}
\end{align*}
Several  inference algorithms are available for this task, e.g. Markov Chain Monte Carlo \cite{jerrum1997markov}, Gibbs sampling  \cite{gelfand2000gibbs}, or no-u-turn Hamiltoninan Monte Carlo (HMC)~\cite{hoffman2014no}.
In recent years, gradient-based inference algorithms such as stochastic variational inference \cite{hoffman2013stochastic} have become popular, because they can scale to larger models by converting an inference problem into an optimization problem.
\item \textbf{Intervention:} Apply the intervention to the SCM to generate a mutilated SCM as in~\figref{fig:mapkNet}(b). 
In the \ac{MAPK} \ac{SCM}, $Mek = f_{Mek}(Raf,N_{Mek})$ is replaced with $Mek=m'$ as shown in~\figref{fig:mapkNet}(b).
\item \textbf{Prediction:}  Generate samples from the mutilated SCM using the estimated posterior distribution over the exogenous variables $\hat{N}_{Raf}$ and $\hat{N}_{Erk}$  to obtain the counterfactual distribution, as shown in~\figref{fig:mapkNet}(b).
\end{enumerate}
\medskip \noindent {\textbf{Causal effects}}
We distinguish between two causal effects. The first is the \acf{ate}, defined as the difference between the outcome of a hypothetical intervention and the observed outcome in the entire population.
In the MAPK example, the ATE of $Erk$ upon an intervention fixing $Raf=r^{\prime}$ is:
\begin{eqnarray}
& \left\{Erk_{do(Raf= r^{\prime})}  - Erk  \right\} &
\end{eqnarray}
This requires no observational data, and therefore the \acf{ate} can be inferred with forward simulation.

On the other hand, the \acf{ite} is defined as the difference between the outcome of a hypothetical intervention and the observed outcome for a specific individual or context.
In the MAPK example, the \acl{ite} of $Erk$ upon an intervention fixing $Raf=r^{\prime}$ in a context where  $Raf = r$, $Mek = m$, $Erk = e$ is:
\begin{eqnarray}
\left\{Erk_{do(Raf= r^{\prime})}  - Erk  \right\} | Raf = r, Mek = m, Erk = e
\end{eqnarray}
The \ac{ite} shares stochastic components of the noise variables between observational and interventional data, and is therefore often more precise than a comparison based on a direct simulation~\cite{ness2019integrating}. 

In cases where domain knowledge is available to describe the systems dynamics in the form of an SDE, the system at equilibrium can be translated into an \ac{SCM} to enable counterfactual reasoning and estimation of the \acl{ite}~\cite{blom2018beyond,ness2019integrating}.
Unfortunately, this process is challenging in novel and poorly studied systems, due to our limited ability to establish the structure of the causal graph.

\medskip \noindent {\textbf{Structured knowledge graphs}}
Although there exist a multitude of biological knowledge bases that are manually curated from the literature~\cite{Cerami2011,Fabregat2018,Kanehisa2017,Perfetto2016,Slenter2018}, the systems biology community has coalesced around a small number of structured knowledge representations that differ mainly in their intended purpose.
For example, the \acf{biopax} was designed for pathway database integration~\cite{demir_2010}, and the \acf{SBGN} was designed for graphical layout~\cite{le2009systems}.

In contrast, the \acf{BEL} was specifically designed for manual extraction and automated integration of author statements about causal relationships  among biological entities, biological processes, and cellular-level observable phenomena~\cite{hoyt2019}.
The syntax of a BEL statement is comprised of a  triple in the form of \{\textit{subject}, \textit{predicate}, \textit{object}\}.
Each subject and object represents an activity or abundance whose entities are grounded using terms from  standard namespaces.
If the subject directly increases the abundance or the activity of the object, we represent this with \verb+=>+, and for directly decreasing relationships, we use \verb+=|+.
BEL statements  can be chained together from the object of the first statement to the subject of the next statement, as shown in~\figref{fig:BEL} for the case of the \ac{MAPK} pathway.

\ac{BEL} provides a number of valuable features for causal modeling.
First, the restriction of \ac{BEL} edges to causal relations implies the topology of the BEL graph can be  reflected in the topology of the causal model.
Second,  the language is expressive enough for humans to manually curate a wide range of biological concepts, but formal enough to serve as a training corpus for \acf{bionlp} competitions~\cite{madan_2019}.
Third, the \ac{BEL} ecosystem is sufficiently mature that causal knowledge represented in other languages can be readily converted to \ac{BEL}~\cite{hoyt_2019,hoyt_2018}.
\begin{center}
    \begin{figure*}
        \begin{small}
            \begin{verbatim}
                kin(p(fplx:RAF)) => kin(p(fplx:MEK))
                kin(p(fplx:MEK)) => kin(p(fplx:ERK))
            \end{verbatim}
        \end{small}
        \caption{\small \textbf{Example BEL statement} The statement details the processes in the MAPK signaling pathway in~\eqref{mapkNet}. The first line states that the kinase activity of RAF directly increases the kinase activity of  MEK. The second line  states that kinase activity of MEK directly increases the kinase activity of ERK. \label{fig:BEL}}
    \end{figure*}
\end{center}    

\section{Methods}
\subsection{Notation, definitions and assumptions}
Let $\mathbf{X}=\{X_i\}$ be a set of variables, such as molecular activities in a signaling pathway.
Let  $\mathbf{P}=\{P_j\}$ be a set of causal predicates that link these variables, such as increases, or regulates.
Using this notation, we define a knowledge graph $\mathbb{K}$ as a set of $k$ triples
\begin{eqnarray}
\mathbb{K} = \{X_i, P_j, X_{i^{\prime}}\ | X_i \in \mathbf{X}, P_j \in \mathbf{P}, X_{i^{\prime}} \in \{\mathbf{X} \setminus X_i \} \}_{j=1}^k
\end{eqnarray}
We define a causal query $\mathbb{Q}$ as a set  $\{\mathbf{X}^{\mathrm{c}},\ \mathbf{X}^{\mathrm{e}},\ \mathbf{X}^{\mathrm{z}}\}$ of variables that are potential causes, effects and covariates of interest for the biological investigation, where
\begin{eqnarray*}
\mathbf{X}^{\mathrm{c}} \subset \mathbf{X},\ \ \mathbf{X}^{\mathrm{e}} \subset \mathbf{X} \backslash \mathbf{X}^{\mathrm{c}},\  \mathrm{and} \ \mathbf{X}^{\mathrm{z}} \subset \mathbf{X} \backslash \mathbf{X}^{\mathrm{c}}\backslash \mathbf{X^{\mathrm{e}}}
\end{eqnarray*}
A pathway $\mathbb{P}(X_1,X_{k^{\prime}+1})$, $k \le k^{\prime}$ is a sequence of a subset of triples from $\mathbb{K}$, where the object of the previous triple is subject of the next triple
\begin{eqnarray}
\left\{\left(X_1, P_1, X_2\right), \left(X_2, P_2, X_3\right), \ldots, \left(X_{k^{\prime}}, P_{k^{\prime}}, X_{k^{\prime}+1}\right)\right\}
\end{eqnarray}

Our goal is to query the knowlege graph to generate a qualitative causal model $\mathbb{B}$ that links the causes, the effects and the covariates of interest.
Importantly, the query result $\mathbb{B}$ induces a directed acyclic graph $G$ with $p$ variables from $\mathbf{X}$ as nodes, and causal relations from $\mathbf{P}$ as edges.

We assume that every variable in $\mathbb{B}$ is continuous.
We denote $\mathbb{D} = \{X_{1j},X_{2j},...,X_{pj}\}_{j=1}^{m}$ the observational data of $m$ samples from the joint distribution $\mathcal{P}(\mathbf{X;\theta})$.
The distribution is specified in terms of parameters $\theta$.
We denote $\mathbf{R} \subset \mathbf{X}$ a set of nodes in $G$ without parents.

\subsection{Querying a knowledge graph to obtain a qualitative causal model}

\begin{algorithm}[t!]
\begin{small}
\caption{
  \small \textit{\acf{query2bel} algorithm}
 \newline {\bf Inputs:} knowledge graph $\mathbb{K}$
 \newline  $~~~~~~~~~~~~$ causal query $\mathbb{Q}=\{\mathbf{X}^{\mathrm{c}},\ \mathbf{X}^{\mathrm{e}},\ \mathbf{X}^{\mathrm{z}}\}$
\newline {\bf Outputs:} $\mathbb{B}$
}
\label{query2bel}
\begin{algorithmic}[1]
  \Procedure{query2bel}{$\mathbf{X}^{\mathrm{c}},\mathbf{X}^{\mathrm{e}},\mathbf{X}^{\mathrm{z}},\mathbb{K}$}
\LineComment{Get all pathways from cause to effect} \label{query_cause_to_effect}
  \For {{\bf each} cause $X_{i}^{\mathrm{c}}\in \mathbf{X}^{\mathrm{c}}$ and for each effect $X_{j}^{\mathrm{e}}\in \mathbf{X}^{\mathrm{e}}$}
    \State{ find all pathways $\left\{\mathbb{P}\left(X_{i}^{\mathrm{c}},X_{j}^{\mathrm{e}}\right)\right\}$ }
    \EndFor
    \LineComment{Get all pathways from covariates to causes} \label{query_covariate_to_cause}
 \For {{\bf each} covariate $X_{i}^{\mathrm{z}}\in \mathbf{X}^{\mathrm{z}}$ and for each cause $X_{j}^{\mathrm{c}}\in \mathbf{X}^{\mathrm{c}}$ }
    \State{ find all pathways $\left\{\mathbb{P}\left(X_{i}^{\mathrm{z}},X_{j}^{\mathrm{c}}\right)\right\}$ } 
 \EndFor
\LineComment{Get all pathways from covariates to effects} \label{query_covariate_to_effect}
 \For {{\bf each} covariate $X_{i}^{\mathrm{z}}\in \mathbf{X}^{\mathrm{z}}$ and for each effect $X_{j}^{\mathrm{e}}\in \mathbf{X}^{\mathrm{e}}$ }
    \State{ find all pathways $\left\{\mathbb{P}\left(X_{i}^{\mathrm{z}},X_{j}^{\mathrm{e}}\right)\right\}$ } 
 \EndFor
 \State{$\mathbb{B} = \left\{\mathbb{P}\left(X_{i}^{\mathrm{c}},X_{j}^{\mathrm{e}}\right)\right\}\cup\left\{\mathbb{P}\left(X_{i}^{\mathrm{z}},X_{j}^{\mathrm{c}}\right)\right\}\cup\left\{\mathbb{P}\left(X_{i}^{\mathrm{z}},X_{j}^{\mathrm{e}}\right)\right\} $}
 \State{{\bf return} $\mathbb{B}$}
\EndProcedure
\end{algorithmic}
\end{small}
\end{algorithm}

Given a biological knowledge graph $\mathbb{K}$ and a causal query of interest $\mathbb{Q}$, our first objective is to generate a qualitative causal model $\mathbb{B}$ capable of answering the query. 
To this end, we need to explore all potential directed acyclic paths in $\mathbb{K}$ from the cause to the effect in $\mathbb{Q}$, and then consider all covariates that may act as confounders of the causal question. 
This is done with the steps in \algref{query2bel}. The algorithm can be implemented on any knowledge graph that represents causal relationships as directed edges, such as \ac{BEL} or the \acl{SBGN} \acl{AF} (\ac{SBGN}-\ac{AF}) language~\cite{mi_2015}.

In the case of MAPK, the qualitative causal model that is capable of answering the counterfactual question in~\eqref{eq:exampleCounterfactual} corresponds to the result of this query: $\mathbb{Q}=\{\mathbf{X}^{\mathrm{c}}=kin(p(\aca{mek})), \mathbf{X}^{\mathrm{e}}=kin(p(\aca{erk})), \mathbf{X}^{\mathrm{z}}=kin(p(\aca{raf}))\}$.

We execute \algref{query2bel} step \ref{query_cause_to_effect} to obtain all pathways from the cause to the effect:
$$kin(p(\aca{mek}))\rightarrow kin(p(\aca{erk}))$$
We execute \algref{query2bel} step \ref{query_covariate_to_cause} to obtain  all pathways from the covariate to the cause:
$$kin(p(\aca{raf}))\rightarrow kin(p(\aca{mek}))$$
We execute \algref{query2bel} step \ref{query_covariate_to_effect}, but since there are no new pathways from the covariate $kin(p(\aca{raf}))$ to the effect $kin(p(\aca{erk}))$, we obtain the empty set.
The final returned model is:

$$kin(p(\aca{raf}))\rightarrow kin(p(\aca{mek}))\rightarrow{kin(p(\aca{erk}))}$$

\subsection{Compiling a qualitative causal model to a quantitative \acl{SCM} }
Our second objective is to express the qualitative causal structure in $\mathbb{B}$ into a quantitative \ac{SCM}, and estimate the parameters of the SCM from experimental data.
These steps are described in Algorithm~\ref{beltoscm}. 

\medskip \noindent {\textbf{Input} The algorithm takes as input a \ac{BEL} causal query result $\mathbb{B}$ and observed measurements on its variables $\mathbb{D}$.

\medskip \noindent {\textbf{Get network structure $G$ from $\mathbb{B}$ (\algref{beltoscm} line \ref{Step:GFromB})}}
Since a set of \ac{BEL} statements identifies parents and children, it induces a causal network structure. We determine this structure by traversing \ac{BEL} statements with the breadth first search approach, starting with root variables (such as $Raf$ in Figure~\ref{fig:BEL}). For all the non-root variables, the algorithm waits until all the parents are traversed. 

\medskip \noindent {\textbf{For each root node $R$, use $\mathbb{D}$ to estimate parameters $\theta$ of $\mathcal{P}(R ; \theta)$(\algref{beltoscm} line \ref{Step:GetThteaForRoot})}}
In order to specify the SCM, we need to define the type and parameters of the marginal probability distributions of the root variables $\mathcal{P}(R ; \theta)$.
The BEL statements provide prior knowledge about the distribution in a parametric form.
Therefore, this step involves techniques such as maximum likelihood to estimate the parameters of this distribution.

\begin{algorithm}[t!]
\begin{small}
\caption{
 \small \textit{\acf{bel2scm} algorithm}
\newline {\bf Inputs:} BEL statements $\mathbb{B}$
\newline $~~~~~~~~~~~~$$\mathbb{D} \sim P(X_1,...,X_p)$
\newline {\bf Outputs:} $SCM$ $\mathbb{M} = \{f_i(\mathbf{PA}_i, N_i)\}_{i=1}^p$
}
\label{beltoscm}
\begin{algorithmic}[1]
\Procedure{bel2scm}{$\mathbb{B}$, $\mathbb{D}$}
\State {$\mathbb{M} = \{\}$}
\State {Get network structure $G$ from $\mathbb{B}$.} \label{Step:GFromB}
 \For {{\bf each} $R \in \mathbf{R}$ in $G$}
        \LineComment{{\footnotesize Use $\mathbb{D}$ to estimate parameters $\theta$ of $\mathcal{P}(R ; \theta)$  }}\label{Step:GetThteaForRoot}
        \State{$~~~~$ $\theta = arg\, max_{\theta} \mathcal{P}(R; \theta \mid \mathbb{D})$}
         \LineComment{{\footnotesize Reparameterize $\mathcal{P}(R; \theta)$ in terms of $f_R$ and $N_R$}}\label{Step:ReparameterizeRoot}
		\State{$~~~~$ $N_{R} \sim \mathcal{N}(0,1)$}
        \State{$~~~~$ $f_R(N_R) = F_{ \mathcal{P}(R;\theta)}^{-1}(N_{R})$ }
        \State{$\mathbb{M}$.Add($f_R(N_R)$)}\label{Step:AddRootToSCM}
 \EndFor
 \For {{\bf each} $X$ $\in \{\mathbf{X} \setminus \mathbf{R}\}$ in $G$ } 
       \LineComment{{\footnotesize Estimate parameters  $\mathbf{w}$ and $b$ of sigmoid function}}\label{Step:ModelTraining}
       \State{$~~~~$ $\log(\frac{X}{\beta_X - X}) = \mathbf{w}^{\prime} \mathbf{PA}_X + b$}
       \LineComment{{\footnotesize Define distribution of $N_X$ from model residuals.}}\label{Step:ModelResiduals}
       \State{$~~~~$ $residual = X - \frac{\beta_X}{1 + \exp(-\mathbf{w}^{\prime} \mathbf{PA}_X - b)}$}
       \State{$~~~~$ $N_X \sim \mathcal{N}(0,MSE(residual)$)}
       \LineComment{{\footnotesize Get $f_X(\mathbf{PA}_X, N_{X})$ with additive $N_X$.}} \label{Step:GetF}
       \State{$~~~~$ $f_{X}(\mathbf{PA}_X, N_{X}) = \frac{\beta_X}{1 + exp(-\mathbf{w}^{\prime}_{X} \mathbf{PA}_X - b_{X})} + N_{X}$}
       \State{$\mathbb{M}$.Add($f_X(\mathbf{PA}_X, N_X)$).}\label{Step:GenerateSCM}
       \EndFor
\State {\textbf{return} $\mathbb{M}$}\label{Step:ReturnM}
\EndProcedure
\end{algorithmic}
\end{small}
\end{algorithm}

For example, in a stochastic MAPK system at equilibrium the root variable the number of active $Raf$ in a cell follows a Binomial distribution. When the maximum number of active or inactive particles in the system is large, the Binomial distribution can be approximated with a Normal distribution with $\theta_{Raf} = (\mu_{Raf}, \sigma^2_{Raf})$. We then estimate $\theta_{Raf}$ using maximum likelihood from the observed $Raf$ in $\mathbb{D}$.

\medskip \noindent {\textbf{For each root node $R$, reparameterize $\mathcal{P}(R; \theta)$ in terms of $f_R$ and $N_R$  (\algref{beltoscm} line \ref{Step:ReparameterizeRoot})}} The specification of an SCM requires us to separate the deterministic and the stochastic components of variation of each variable as shown in~\figref{fig:mapkNet}. We accomplish this using a reparameterization technique popularized by variational autoencoders \cite{rezende2014stochastic}, which was shown to make counterfactual inference consistent with core biological assumptions \cite{NIPS2019_9569}. In the case of root nodes, we reparameterize $\mathcal{P}(R;\theta)$ with Uniform(0,1), and then pass it to the inverse CDF of $\mathcal{P}(R;\theta)$, as follows

\begin{eqnarray}
        \mathrm{Original:}\ && R \sim \mathcal{P}(R;\theta) \\
        \mathrm{Reparametrized:}\ && N_{R} \sim \text{Uniform}(0,1) \nonumber \\
        && f_R(N_R) = F_{ \mathcal{P}(R;\theta)}^{-1}(N_{R}) \nonumber 
\label{reparameterization}
\end{eqnarray}
where $F_{ \mathcal{P}(R;\theta)}^{-1}(N_{R})$ is the inverse cumulative distribution function of $\mathcal{P}(R;\theta)$.
In the case of MAPK, since $Raf$ follows a Normal distribution with parameters $\theta_{Raf}$, the reparameterization simplifies even further to
\begin{eqnarray}
        \mathrm{Original:}\ && Raf \sim \mathcal{N}(\mu_{Raf},\sigma^2_{Raf}) \label{Eqn:Raf} \\
        \mathrm{Reparametrized:}\ && N_{Raf} \sim \mathcal{N}(0,1) \nonumber \\
        && f_{Raf}(N_{Raf}) = \sigma_{Raf} N_{Raf} + \mu_{Raf} \nonumber
\label{reparameterization}
\end{eqnarray}

\begin{figure}[t!]
\begin{center}
\includegraphics[width=0.3\textwidth]{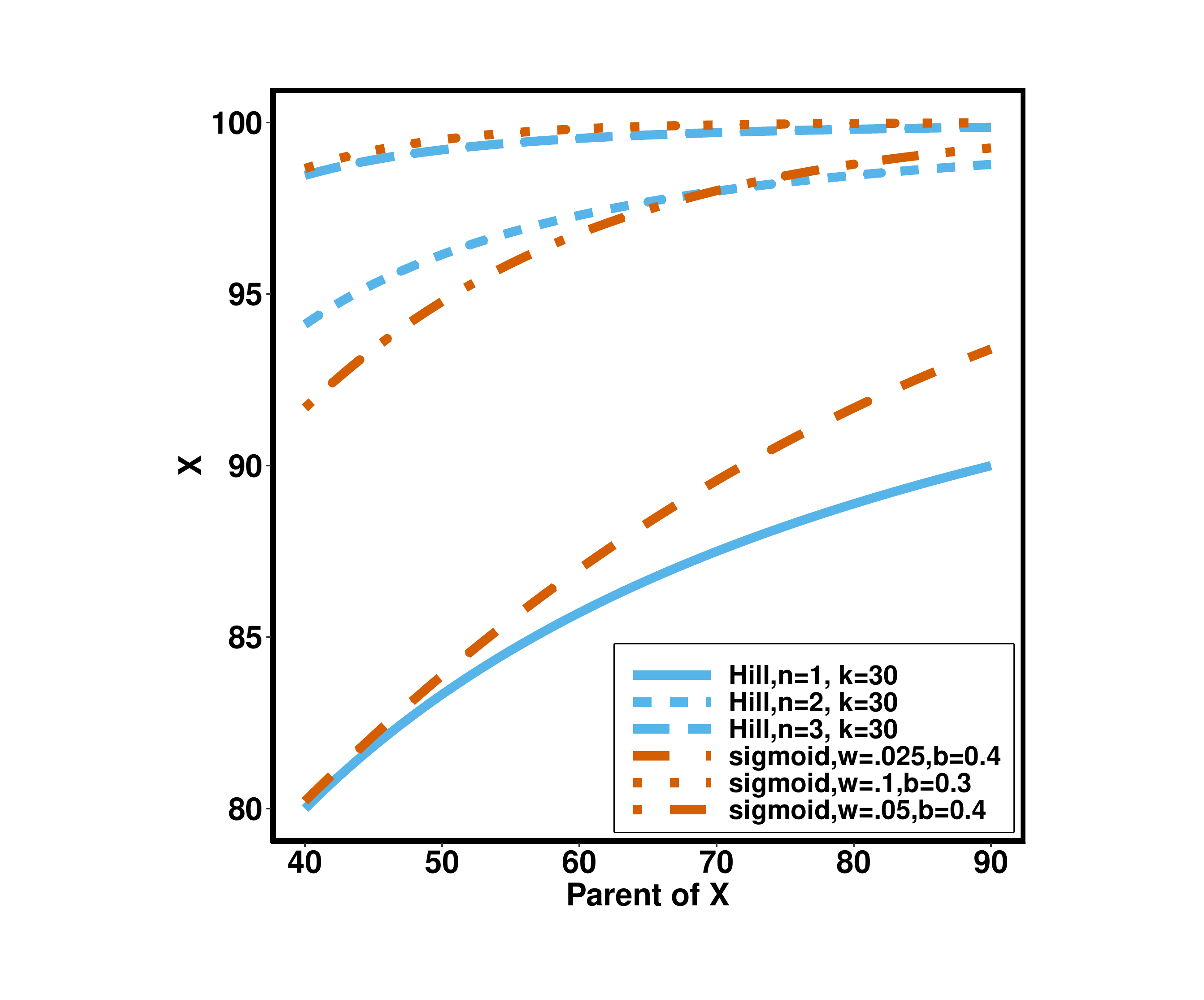}
\end{center}
\vspace{-.15in}
\caption{\small   \textbf{Examples of Hill function and sigmoid function for two variables}  $X$ is a single node that has a single parent  $ \mathbf{PA}_X$. We use the Hill function ($X = \beta \frac{\mathbf{PA}_X^n}{K^n + \mathbf{PA}_X^n}$) and sigmoid function as in  \eqref{eqn:sigmoidFunc} to predict the value of $X$ given its parent value. In the Hill function, $K$ is the activation rate, $n$ defines the steepness of function and $\beta$ is fixed at 100. Blue lines correspond to Hill equation  with $K = 30$ and  $n \in \{1, 2, 3\}$. Brown lines correspond to sigmoid function where $b \in\{0.4,0.3,0.4\}$ and $w \in \{0.025,0.1, 0.5\}$}  \label{fig:hillSigmoid}
\end{figure}

\medskip \noindent {\textbf{ Add $R$ to $\mathbb{M}$ (\algref{beltoscm} line \ref{Step:AddRootToSCM})}}
For each root node, we add the corresponding function $f_R(N_R)$ and its noise variable $N_R$ to $\mathbb{M}$.
For example, since MAPK has only one root node $Raf$, the Algorithm adds $f_{Raf}(N_{Raf})$ to $\mathbb{M}$.

\medskip \noindent {\textbf{For each $X \in \{\mathbf{X} \setminus \mathbf{R}\}$, estimate parameters  $\mathbf{w}$ and $b$ of sigmoid function  (\algref{beltoscm} line \ref{Step:ModelTraining})}}
In order to specify the SCM for non-root nodes, we need to define the form (polynomial, linear, non-linear, sigmoid, etc.) of functional assignments linking the measurements on the parent nodes to the measurements on the child.
We chose the functional assignment in the form of a sigmoid function
\begin{equation}
\label{eqn:sigmoidFunc}
\log\left(\frac{X}{\beta_X - X}\right) = \mathbf{w}^{\prime}  \mathbf{PA}_X + b
\end{equation}
where $\beta_X$ is the maximum number of activated protein molecules.  For a node $X$ with $q$ parents, $\mathbf{PA}_X$ is a $q \times 1$ vector of measurements on the parent nodes, $\mathbf{w}$ is a $1 \times q$ vector of weights, $\mathbf{w}^{\prime}$ is the transpose of $\mathbf{w}$, and $b$ is a scalar bias.   
Parameters $\mathbf{w}$ and $b$ of the sigmoid function are estimated from the data, e.g. using smooth $L_1$ loss function.

In the example of the \ac{MAPK} pathway, $f_{Mek}$ has only one parent. Therefore $f_{Mek}$ has the form
\begin{equation}
f_{Mek}(Raf, N_{Mek}) = \frac{\beta_{Mek}}{1 + exp(-{w}_{Mek} Raf - b)} + N_{Mek}
\end{equation}
We use the sigmoid function in~\eqref{eqn:sigmoidFunc} as a special case of the Hill equation.
The full parametric description of the Hill equation has a nuanced precise biochemical interpretation.
For example, the parameter $n$ represents the number of times a protein must be phosphorylated before it becomes active and can therefore be obtained from domain knowledge.
However, it is difficult to estimate this parameter from data.
The sigmoid function maintains the Hill equation's functions, but with a reduced set of parameters that are easier to estimate.
\figref{fig:hillSigmoid} shows that the approximation is reasonable for a range of parameter values. 

\medskip \noindent {\textbf{Define distribution of $N_X$ from model residuals (\algref{beltoscm} line \ref{Step:ModelResiduals})}}
Similarly to the root variables, for non-root variables we assume that the noise variables follow Normal distribution with 0 mean. The variance of this distribution is estimated from the residuals of the model fit in the previous step.
For example, in the \ac{MAPK} pathway, $f_{Mek}$ has only one parent $Raf$. Therefore, the residuals of the sigmoid curve fit for $Mek$ are defined as
\begin{equation}
residual_{Mek} = Mek - \frac{\beta_{Mek}}{1 + \exp(-w_{Mek} Raf - b)}
\end{equation}
and the distribution of the noise variable is defined as $N_{Mek} \sim \mathcal{N}(0, MSE(residual_{Mek}))$

\medskip \noindent {\textbf{Get $f_X(\mathbf{PA}_X, N_{X})$ with additive $N_X$ (\algref{beltoscm} line \ref{Step:GetF})}}
The step combines the sigmoid functional assignment and the independent noise variable. 
In the example of $Mek$ in the MAPK pathway, the step outputs
\begin{equation}
f_{Mek}(Raf, N_{Mek}) = \frac{\beta_{Mek}}{1 + exp(-{w}_{Mek} Raf - b)} + N_{Mek}
\end{equation}

\medskip \noindent {\textbf{Add $f_X(\mathbf{PA}_X, N_X)$ to \ac{SCM} (\algref{beltoscm} line \ref{Step:GenerateSCM})}}
The step iteratively adds $(f_X, N_X)$ for all $X \in \mathbf{X}$. 

\medskip \noindent {\textbf{Output (\algref{beltoscm} line \ref{Step:ReturnM})}} The algorithm returns a generative structural causal model $\mathbb{M} = \{f_i(\mathbf{PA}_i, N_i)\}_{i=1}^p$ where $\mathbf{PA}_i \subset \mathbf{X}$.
For example, in the case of the MAPK model, it returns $[N_{Raf}, N_{Mek}, N_{Erk}, f_{Raf}(N_{Raf}),f_{Mek}(Raf, N_{Mek}), \\f_{Erk}(Mek, N_{Erk})]$.

\subsection{Counterfactual inference procedure}
The generated \ac{SCM} enables counterfactual inference using a standard procedure~\cite{pearl2009causal}.
Given a new observation $\mathbb{D}^{new}$,
\begin{enumerate}
\item \textbf{Abduction:} Update the probability $P(N_X)$ to obtain $P(N_X|\mathbb{D}^{new})$.
\item \textbf{Action:} Replace the equations determining the variables in set $\mathbf{X}^{\textrm{c}}$ by $\mathbf{X}^{\textrm{c}} = \mathbf{x}^{\textrm{c}\prime}$.
\item \textbf{Prediction:} Sample from the modified model to generate the target distribution $\mathbf{X}^{\textrm{e}}_{do(\mathbf{X}^{\textrm{c}} = \mathbf{x}^{\textrm{c}\prime})}$.
\end{enumerate}
After generating the target distribution of the intervention model, we estimate causal effects.
~\algref{ce}  describes the detailed steps of both counterfactual inference (with $\mathbb{D}^{new}$) and forward simulation (if  $\mathbb{D}^{new}$ is empty)

\begin{algorithm}[h!]
\begin{small}
\caption{
 \small \textit{Estimate causal effect on $X^E$ upon intervening on $X^C$}
\newline {\bf Inputs:} New data point $\mathbb{D}^{new}$\\
$~~~~~~~~~~~~$ effect node $X^E$\\
$~~~~~~~~~~~~$ observational data for effect node $\mathbb{D}^E \in \mathbb{D}^{new}$\\
$~~~~~~~~~~~~$ intervention value $c$\\
$~~~~~~~~~~~~$ node to intervene upon $X^C$\\
$~~~~~~~~~~~~$ number of iteration $I$\\
$~~~~~~~~~~~~$ network structure $G$\\
$~~~~~~~~~~~~$ SCM $\mathbb{M}$
\newline {\bf Outputs:} Causal Effect $CE$}
\label{ce}
\begin{algorithmic}[1]
\Procedure{getCausalEffect}{$\mathbb{D}^{new}, E, \mathbb{D}^E, X^C , c, I, G, \mathbb{M}$}
\State{$\hat{N} = \{\}$}
 \LineComment{{\footnotesize Interventional data for effect node $X^E$}}
\State{$\mathbb{ID}^E = \{\}$}
\For{$I$}
 \For {{\bf each} $X$ $\in \{\mathbf{X} \setminus X^C\}$ in $G$} 
 \LineComment{{\footnotesize \textbf{Abduction:} Apply stochastic variational inference}}
\State{$~~~~$ $\hat{N}_X = SVI(\mathbb{D}^{new}) $ }
\State{${\hat{N}}$.Add($\hat{N}_X$)}

\EndFor
 \LineComment{{\footnotesize \textbf{Action:} Apply intervention on $X^C$}}
\State{$~~~~$ $CM = pyro.do(\mathbb{M}, {X^C = c})$}
 \LineComment{{\footnotesize Get posterior of $CM$ with importance sampling}}
\State{$~~~~$ $CMP= pyro.infer.Importance(CM, \hat{N})$}
 \LineComment{{\footnotesize \textbf{Prediction:} Get EmpiricalMarginal (EM) for $X^E$}}
\State{$~~~~$ $CMM = pyro.infer.EM(CMP, X^E)$}
\State{$\mathbb{ID}^E$.Add($CMM$)}
\EndFor
\State{$CE = \mathbb{ID}^E - \mathbb{D}^E$}
\State {\textbf{return} $CE$}
\EndProcedure
\end{algorithmic}
\end{small}
\end{algorithm}

\subsection{Implementation}
\ac{query2bel} was implemented manually using a publicly available instance of BioDati Studio~\cite{biodati}, then validated using \ac{INDRA}'s interactive dialogue system Bob with BioAgents~\cite{gyori2017word}.
Parameter estimation in \ac{bel2scm} was implemented in PyTorch. 
Let $C$ be the number of nodes in causal graph $G$ with parents. Let $k$ be the number of iterations for gradient descent, let $N$ be the number of samples in data, and let $d$ be the maximum number of parents in graph $G$.
Computational complexity of parameter estimation step is given by $O(CkNd)$.

SCM-based counterfactual inference was performed with Pyro~\cite{bingham2018pyro}, due to its ability to perform interventions on probabilistic models and scalability to larger models, as described in~\algref{ce}.
Specifically, the implementation relies on the following functionalities in Pyro.
The {\tt pyro.do} method is an implementation of Pearl's do-operator used for causal inference.
The {\tt pyro.infer.SVI} method performs abduction using stochastic variational inference with ELBO loss.
The {\tt pyro.infer.Importance} method performs posterior inference by importance sampling.
The {\tt pyro.infer.EmpiricalMarginal} method performs empirical marginal distribution from the trace posterior’s model.

Experiments in this manuscript took between 13 to 82 seconds depending on the graph size on a system with Intel Core i7 8th Gen CPU, 16 GB RAM and Ubuntu 18.04 Operating System.
The code is available at \aca{bel2scm}.


\section{Case Studies}

Below we introduce two biological case studies investigated using the approach proposed in this manuscript.
The first case study allows us to evaluate the accuracy of the results based on known ground truth.
The second uses counterfactual reasoning to pinpoint the mechanism by which \ac{sars2} infection can lead to a \ac{ards} in severely ill \ac{covid19} patients.
The details of the case studies, parameter values of the simulations, and of the results are at \aca{bel2scm}.

\subsection{Case study 1: the IGF signaling system}
\label{caseStudy1}

\noindent {\textbf{The system}} 
The \ac{igf} signaling pathway (Figure~\ref{fig:igf_ce}) regulates growth and energy metabolism of a cell.
The IGF system has been extensively investigated, and its dynamics are well characterized in form of ODE and SDE models \cite{bianconi2012computational}.  
Activated by external stimuli, insulin-like growth factor (IGF) or epidermal growth factor (EGF)
triggers a signaling event, which includes the MAPK signaling pathway in~\eqref{mapkNet}.
Similarly to~\eqref{mapkNet}, nodes in the system are kinase activities, and edges represent whether the kinase activity of the upstream protein directly increases or decreases the kinase activity of the downstream protein.
However, the system is larger and more complex.
It includes two different paths from $Ras$ to $Erk$, one direct and the other through $PI3K$ and $Akt$.
This challenges estimates of outcomes of interventions.
In this case study, we assume that the \ac{igf} system has no unobserved confounders.

\begin{figure}[t!]
\begin{center}
\includegraphics[width=.24\textwidth]{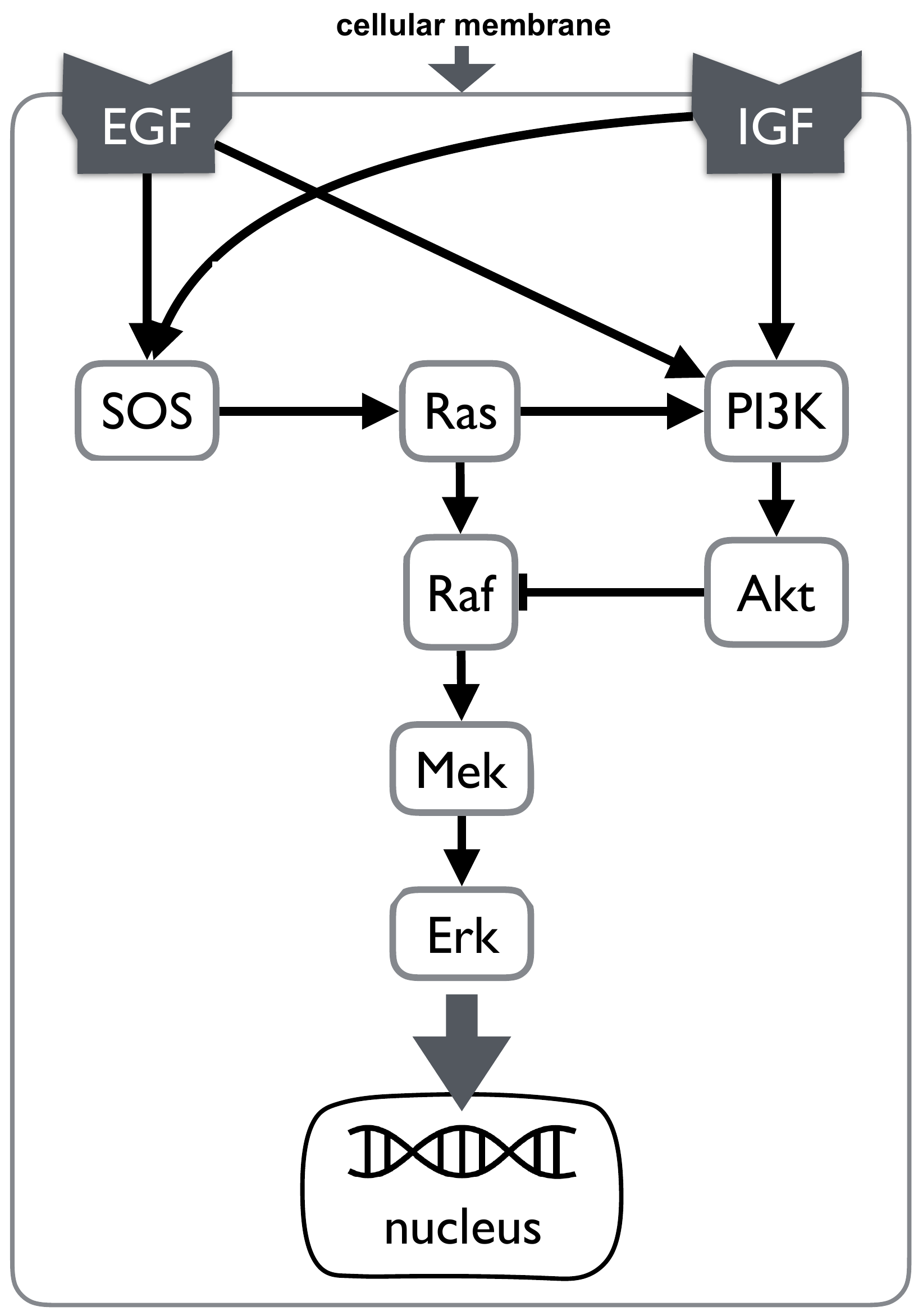}
\end{center}
\caption{\small {\bf Case Study 1: the IGF signaling system} 
The insulin-like growth factor (IGF) and epidermal growth factor (EGF) are receptors of external stimuli, triggering downstream signaling pathways that include the MAPK pathway.
All the relationships between abundances of activated proteins in this network are of the type \textit{increase}, except for the relationship between $Akt$ and $Raf$ which is of the type \textit{decrease}. }  \label{fig:igf_ce}
\end{figure}

\medskip \noindent {\textbf{Intervention}}
We considered two interventions.
The first fixes the kinase activity of $Mek$ to 40.
The second fixes the kinase activity of $Ras$ to 30.

\medskip \noindent {\textbf{Causal effects of interest}}
We are interested in two causal questions.
First,  \textit{ what would have been the kinase activity of $Erk$ had we intervened to fix the kinase activity of $Mek$ to 40?}
The second query is as above, but with the intervention fixing the kinase activity of $Ras$ to 30.
More formally, we are interested in the average treatment effect
\begin{align}
\left\{Erk_{do(Mek= 40)}  - Erk  \right\}  \\
\left\{Erk_{do(Ras = 30)} - Erk  \right\} 
\end{align}
Next, we introduce a new piece of information about a specific data point  generated from the ODE-based simulation.
We wish to estimate the causal effect of intervention for this specific data point.
More formally, we are interested in the individual treatment effect
\begin{align}
\left\{Erk_{do(Mek= 40)}  - Erk  \right\} | \mathbb{D}^{new} \\
\left\{Erk_{do(Ras = 30)} - Erk  \right\} |  \mathbb{D}^{new}
\end{align}
where $\mathbb{D}^{new}$ is a new data point.
We note that this counterfactual inference can only be performed with an SCM.
We wish to compare these estimates of causal effects, in order to characterize the ability of counterfactual inference via $D^{new}$ to improve the precision of the estimates.

\medskip \noindent {\textbf{Evaluation}}
The kinetic equations described by the ODE and SDE represent the true underlying dynamics of the IGF signaling pathway.
Since the ODE and the SDE can estimate the causal effects by forward simulation, we view the estimates as the ground truth.
We then wish to compare the estimates from the SCM against the ground-truth estimates from the ODE and the SDE.
Since an SCM represents causal relationships at steady state, we train the parameters of the SCM using data generated from the ground-truth SDE after it has reached steady state.

We consider two types of evaluations.
First, we compare the estimates of the forward simulation of the ODE and SDE with the forward simulation of the \ac{SCM}.
This allows us to characterize the impact of SCM specification and estimates of weights on the accuracy of causal effects.
We do not expect to see a substantial difference between these two approaches for a correctly specified SCM.
We then compare the \ac{SCM}-based counterfactual inference of causal effects with the estimates based on forward simulation.
We expect that the counterfactual inference will provide more precise estimates, illustrating the statistical efficiency of counterfactual inference as compared to the forward simulation.


\subsection{Case study 2: host response to viral infection}

\medskip \noindent {\textbf{The system}}
Retrospective studies have indicated that high levels of pro-inflammatory cytokine \acf{il6} are strongly associated with severely ill \ac{covid19} patients~\cite{ulhaq_2020}.
One recently proposed explanation for this is the viral induction of a positive feedback loop, known as \acf{il6amp}~\cite{hirano_2020}.
\ac{il6amp} is stimulated by simultaneous activation of \ac{nfkb} and \ac{stat3}~\cite{murakami_2012}. This in turn induces various pro-inflammatory cytokines and chemokines, including \acl{il6}, which recruit activated T cells and macrophages.
This strengthens the \acl{il6amp} into a positive feedback loop leading to a \ac{ards}~\cite{ogura2008interleukin}, which is believed to be responsible for the tissue damage observed in patients with acute respiratory distress syndrome (ARDS)~\cite{hirano_2020}. 

\medskip \noindent {\textbf{Intervention}}
Originally developed to treat autoimmune disorders such as rheumatoid arthritis~\cite{oldfield_2009},
\acf{toci} is an immunosuppressive drug consisting of a recombinant monoclonal antibody that targets the \acl{sil6r} and can effectively block the \ac{il6} signal transduction pathway~\cite{zhang_2020}.
\acl{toci} has emerged as a promising drug repurposing candidate to reduce mortality in severely ill \ac{covid19} patients~\cite{coomes_2020,xaoling_2020}.

\medskip \noindent {\textbf{Causal effect of interest}}
We define a severely ill \ac{covid19} patient as someone with $\aca{ards}>65$.
We are interested in the individual treatment effect (ITE)
\begin{eqnarray}
\label{covCounterfactualQuerry}
& \left\{\aca{ards}_{do({\ac{toci} = 0)}} - \aca{ards}   \right\} | \mathbb{D}^{new} &
\end{eqnarray}
where $\mathbb{D}^{new}$ is an observed patient who received \acl{toci} treatment and became severely ill.
We wish to characterize the severity of \ac{ards} which would have occurred had she not received the treatment.
We further wish to compare the \ac{ite} with the \acf{ate}
\begin{eqnarray}
\label{covATE}
& \left\{\aca{ards}_{do({\ac{toci} = 0)}} - \aca{ards}   \right\}  &
\end{eqnarray}

\medskip \noindent {\textbf{Evaluation}}
\acl{toci} is known to have a strong inhibitory effect on \acl{sil6r}. 
We therefore expect that the severity of the \ac{ards} would have been worse had the patient not received treatment.
Unfortunately, at the time of writing, there were no ODE or SDE-based models of the pathway, nor were there publicly available \ac{covid19} datasets quantifying the kinase activity of the \acl{il6amp} pathway at the single-cell level.
Therefore, we simulated data from a ``ground-truth'' \acl{sigscm}, where the topology reflects the causal structure of the pathway, and the numeric values of the parameters were fixed to reflect our prior qualitative knowledge of the \ac{il6amp} pathway.

We evaluate the \ac{ite} the proposed approach in two ways.
First, we train the parameters of the \ac{SCM} using the simulated data, and compare the counterfactual inference of the \ac{ite} obtained from the ``trained'' \ac{SCM} to  the counterfactual inference of the \ac{ite} from the ``ground-truth'' \ac{SCM}.
This comparison allows us to characterize the impact of weight estimation on the accuracy of causal effects.
We expect that the need to estimate the weights will inflate the variance of the estimates.
Second, we compare the estimates of \ac{ite} to the estimates of the \ac{ate} using the trained \ac{SCM}.
This comparison allows us to characterize the statistical efficiency of counterfactual inference when estimating causal effects.
We expect that the \ac{ite} will provide much more precise estimates.

\section{Results}
\begin{figure}[t!]
\begin{center}
\includegraphics[width=0.23\textwidth]{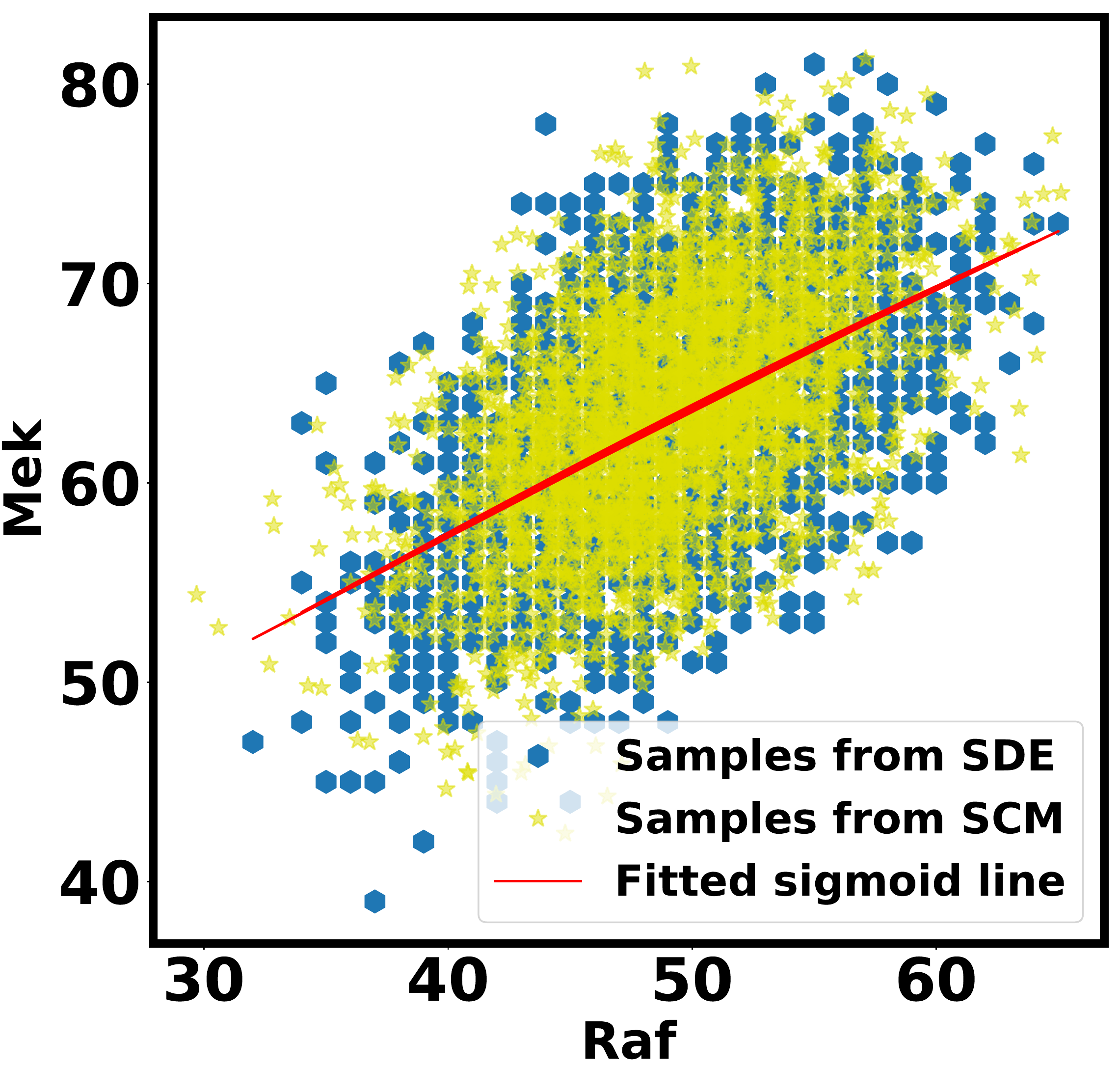}
\end{center}
\vspace{-.15in}
    \caption{\small   {\bf Case study 1: IGF model}  Scatter plot of $Mek$ versus $Raf$. Blue points are the data points generated by SDE. Yellow points are the estimates from \ac{SCM}. The red line is the fitted sigmoid curve in~\algref{beltoscm} line~\ref{Step:ModelTraining}.}  \label{fig:MekRafScatter}
\end{figure}
\subsection{Case study 1: the IGF signaling system}

\begin{figure}[t!]
\begin{center}
\begin{tabular}{ccc}
\includegraphics[width=0.22\textwidth]{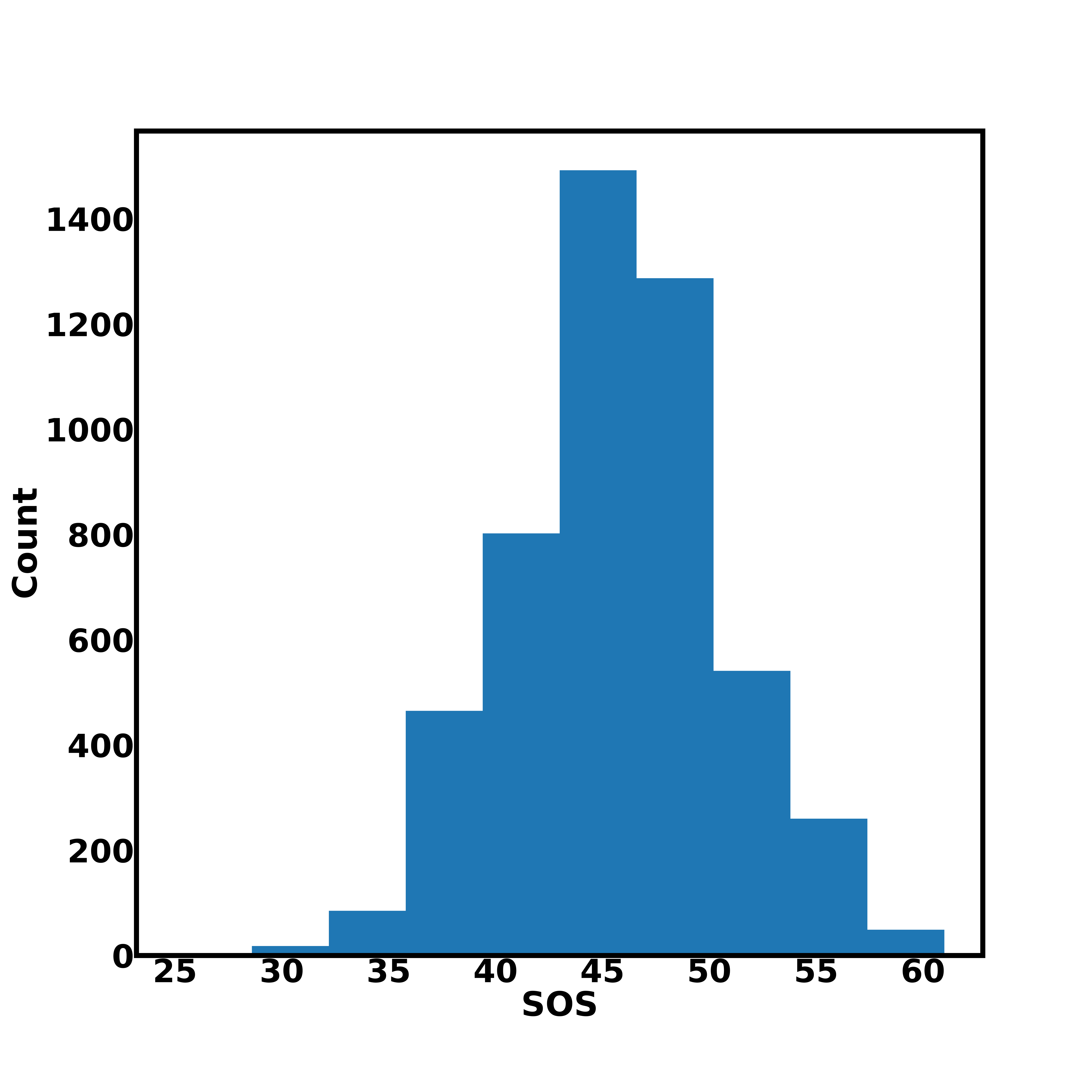} \label{fig:SosHist}&
\includegraphics[width=0.22\textwidth]{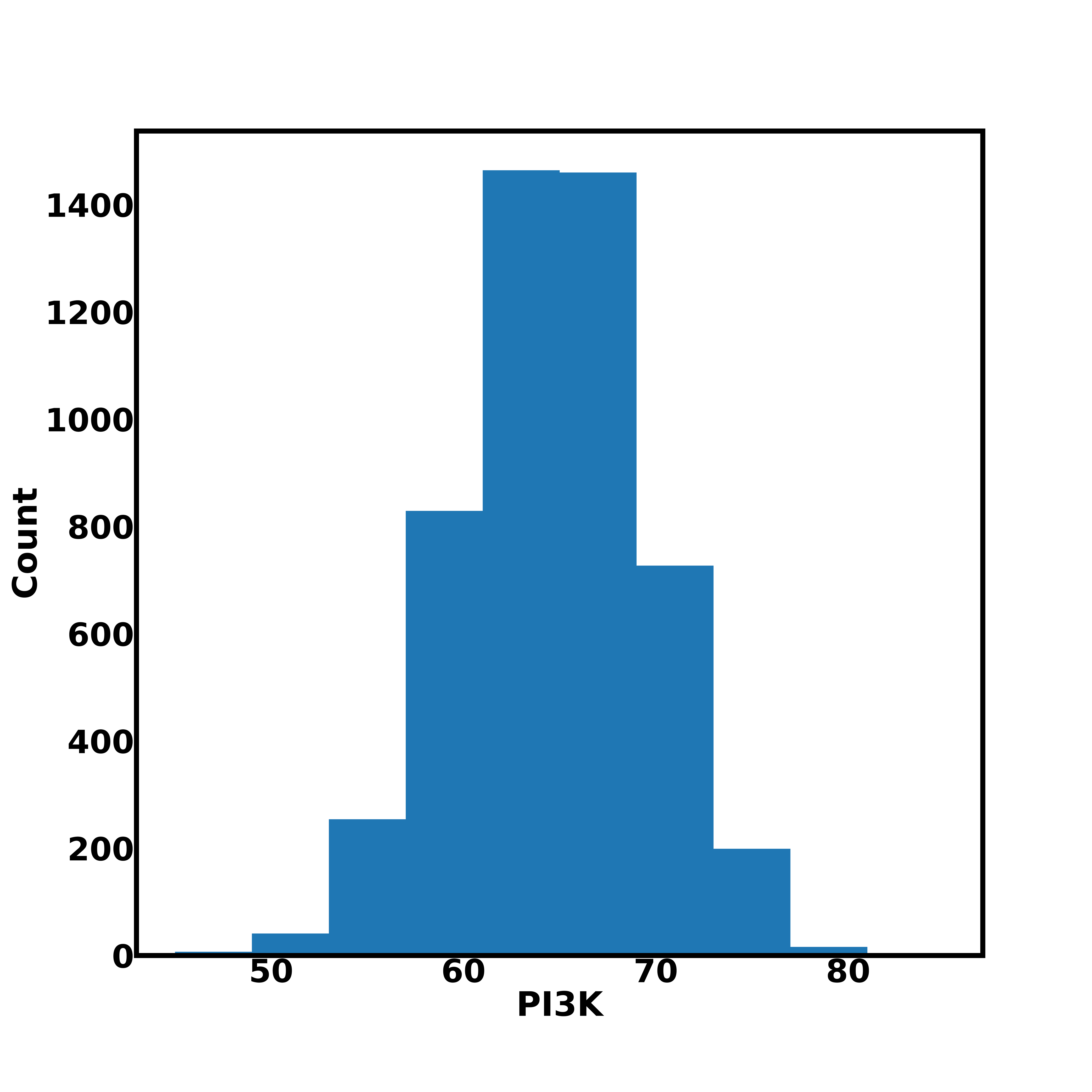}\label{fig:Pi3kHist} \\
(a) & (b)
\end{tabular}
\end{center}
    \caption{\small \textbf{Case study 1: probability distributions of the root nodes of IGF model} (a) Histogram of $SOS$ generated from SDE simulation  (b) As in (a), for $PI3K$. }  \label{fig:RafMekHist_ce} 
\end{figure}

\medskip \noindent {\textbf{Generating BEL causal model}}
The \ac{BEL} representation of the \ac{igf} system was manually curated using BioDati studio~\cite{biodati}, to match the existing ODE and SDE.
The \ac{BEL} representation of the \ac{igf} system specified all the node types as in category \textit{abundance}.
All the relationships between parents and children nodes were of type \textit{increase}, except for the parent node $Akt$, where the relationship was of type \textit{decrease}.

\medskip \noindent {\textbf{Observational data}}
We mimicked the process of collecting observational data by simulating kinase activity from the corresponding \ac{ODE} and \ac{SDE}.
The initial number of particles for the receptor was 37 for $EGF$ and 5 for $IGF$.
The deterministic simulation numerically solved the ODE using the \textit{deSolve}~\cite{soetaert2010solving} R package.
The stochastic simulation used the Gillespie algorithm~\cite{gillespie1977exact} from the \textit{smfsb}~\cite{wilkinson2018package} R package.

\medskip \noindent {\textbf{Appropriateness of model assumptions}} 
SCM-based estimates of functional assignments with sigmoid approximations were well within the range of the SDE-based data (as shown for $Raf$ and $Mek$ in \figref{fig:MekRafScatter}).
Similar results were obtained for estimates of $Ras$, $PI3K$, $AKT$, $Raf$, and $Erk$. The fitted functional assignment had little curvature. This indicates that a more complicated function with more parameters, such as Hill equation, was unnecessary in this case.

To further evaluate the plausibility of the assumptions, \figref{fig:RafMekHist_ce} shows the histograms of the SDE-generated abundances of root nodes, which were not affected by functional assignments in SCM. The shape of the histograms indicate that the assumption of Normal distribution was plausible.

\medskip \noindent {\textbf{Accuracy of causal effects}}
\figref{fig:igfSCMvsSDE}(c) and (d) show that the \acfp{ate} on $Erk$ of fixing $Mek$ and $Raf$, based on forward simulation of ODE, SDE and SCM, were consistent.
\figref{fig:igfSCMvsSDE}(a) and (b) show that the \acfp{ite} based on counterfactual inference has a smaller variance than the \ac{ate}.
Since counterfactual inference reduces nuisance variation by sharing stochastic components in contexts with and without intervention, it increases the statistical efficiency of the estimation.

The \acl{ite} on $Erk$ by fixing $Mek$ was much stronger than the \ac{ite} on $Erk$ by fixing $Ras$ for the following reason. 
While $Mek$ directly influences $Erk$ (i.e., there is a single path from $Mek$ to $Erk$), $Ras$ has two pathways to $Erk$. 
The path through $AKT$ has an inhibiting (deactivation) effect on $Raf$, and estimated negative weights in the sigmoid function in \eqref{eqn:sigmoidFunc}. 
The alternative path, a cascade from $Ras$ to $Erk$, has the opposite (activating) effect on $Erk$. 
The two paths mitigate the overall causal effect of $Ras$ on $Erk$. 

\begin{figure}[t!]
\begin{center}
\begin{tabular}{cc}
\includegraphics[width=0.22\textwidth]{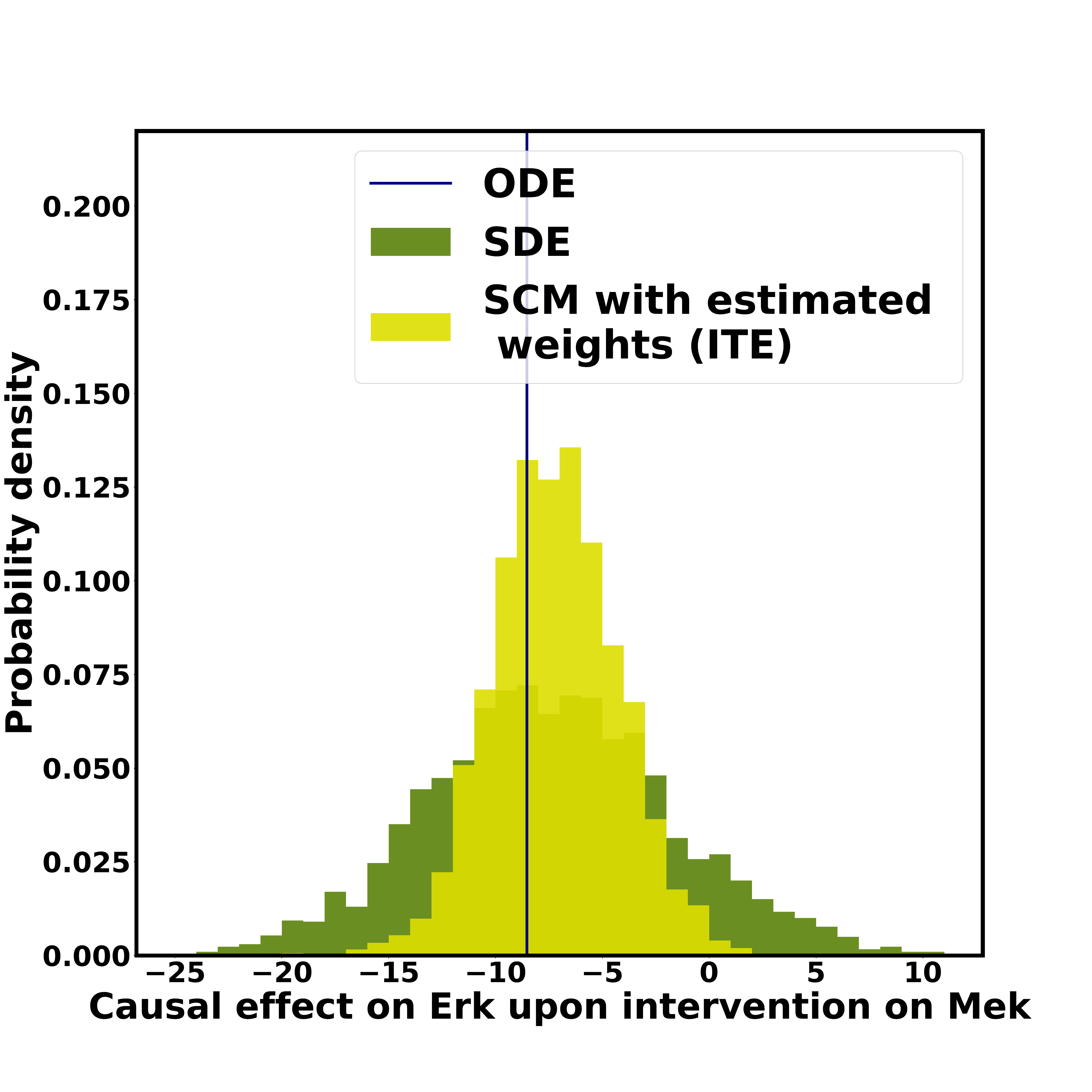}\label{fig:MapkCounterfactual}&
\includegraphics[width=0.22\textwidth]{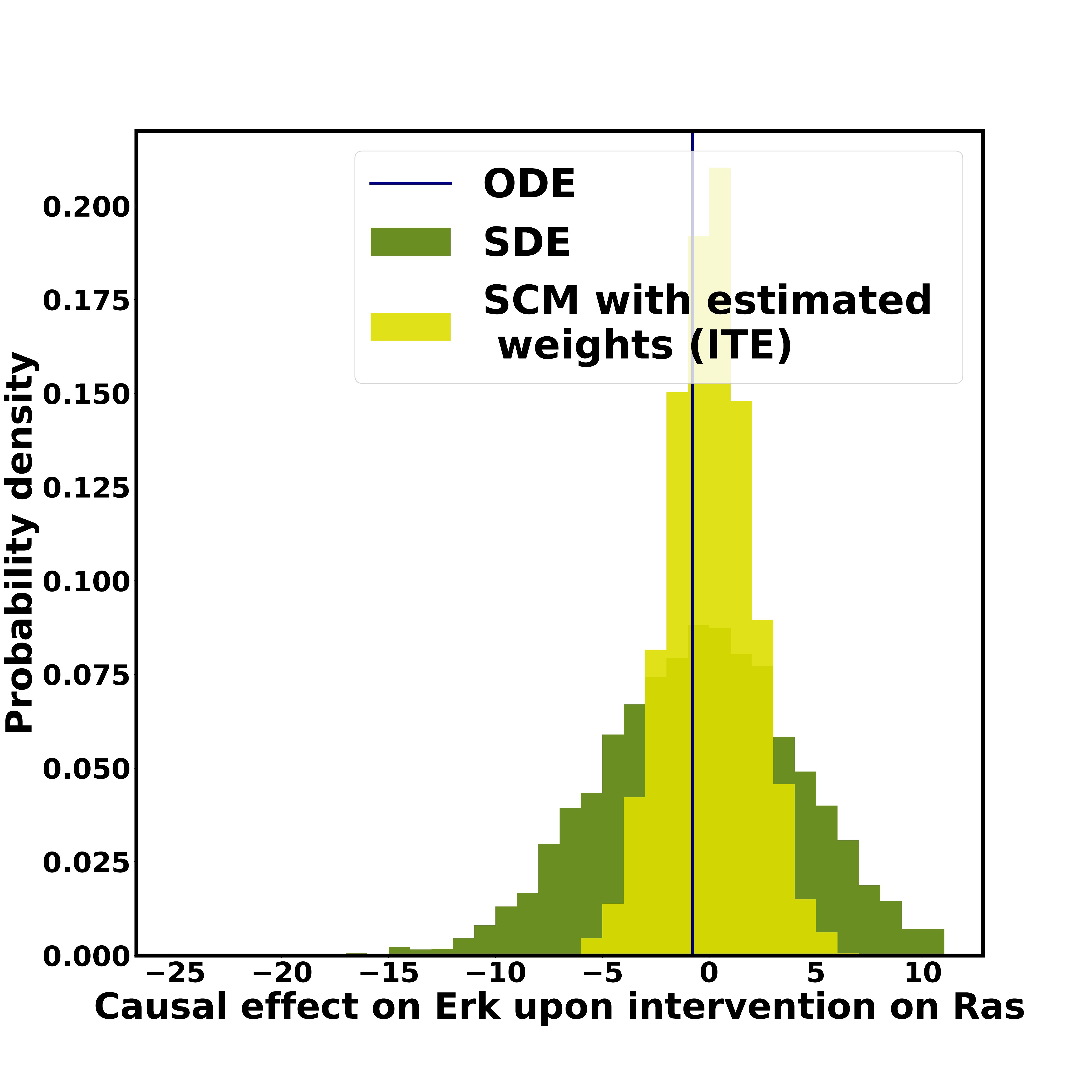}\label{fig:IgfCounterfactual}\\
(a) & (b)\\
\includegraphics[width=0.22\textwidth]{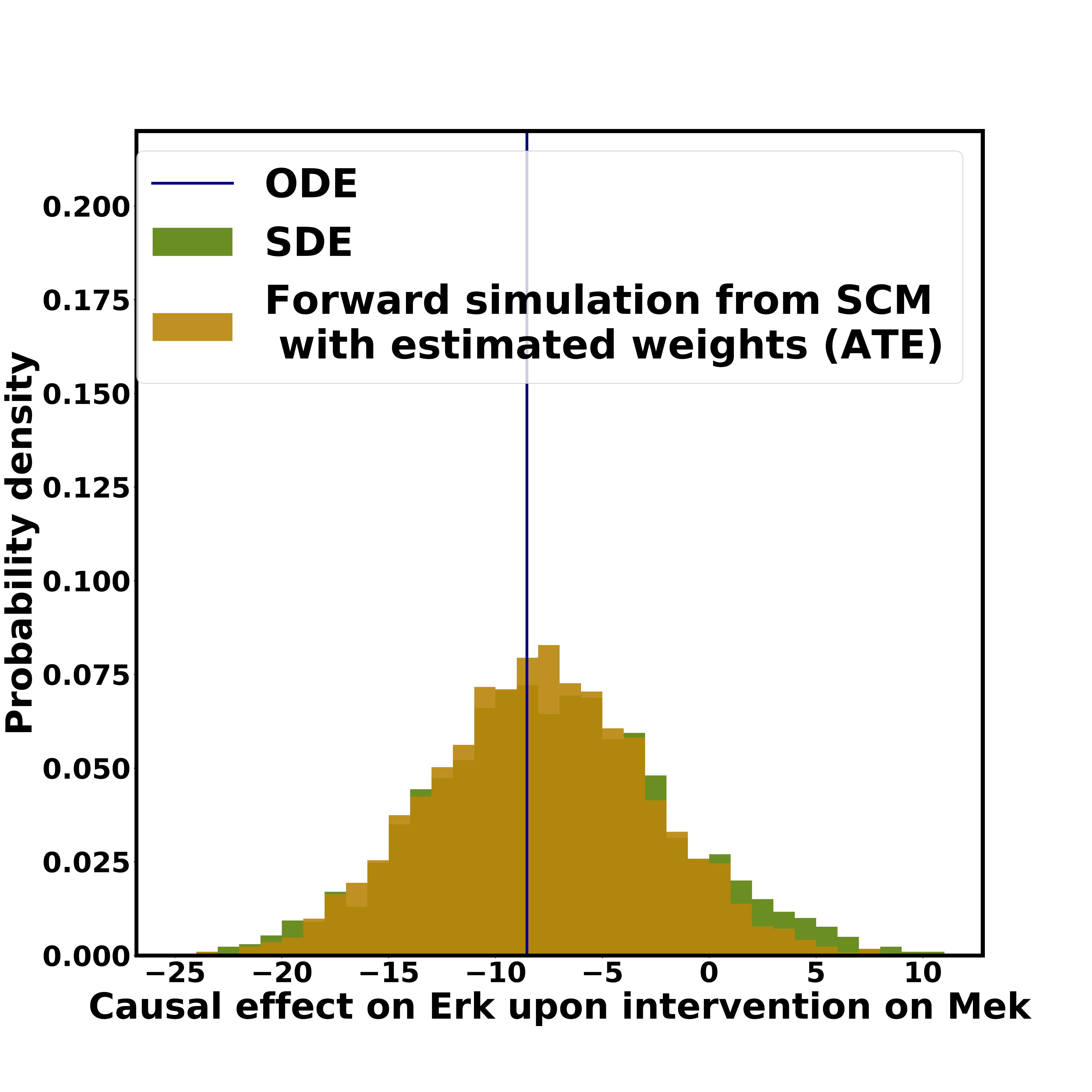}&
\includegraphics[width=0.22\textwidth]{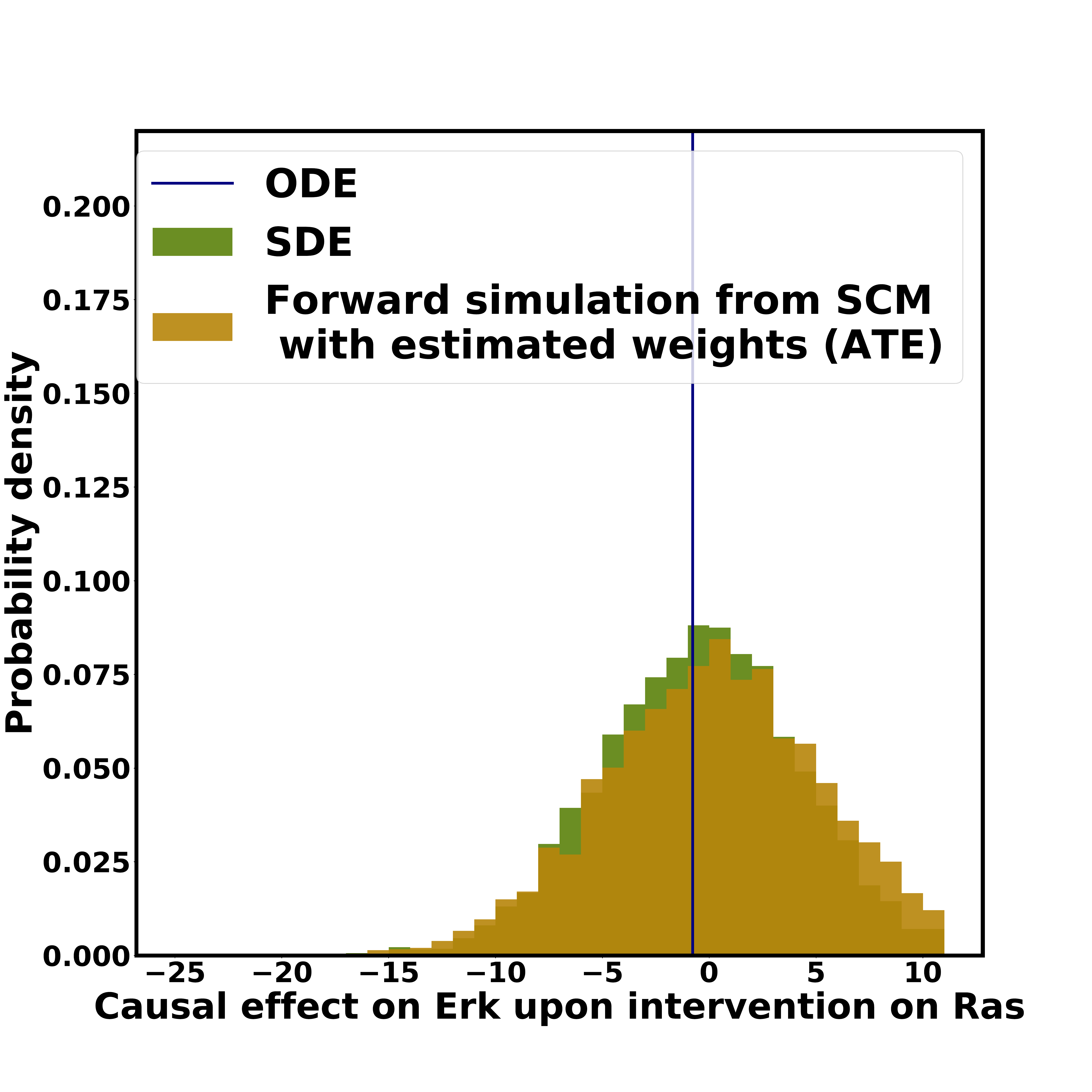}\\
(c) & (d)\\
\end{tabular}
\end{center}
\caption{\small \textbf{Case study 1: estimated causal effects of the IGF signaling pathway using \algref{ce}.}  The ODE and SDE represent the true underlying dynamics of the IGF signaling pathway. The ODE and SDE-based forward simulation can only estimate the average treatment effect. These estimates are viewed as ground truth. In contrast, an \ac{SCM} can estimate both the average treatment effect (ATE) and the individual treatment effect (ITE). (a) Comparison of ITE vs ATE for $Erk$ when $Mek$ is fixed. (b) Comparison of ITE vs ATE for $Erk$ when $Ras$ is fixed. (c) Comparison of SCM, SDE and ODE estimates of the ATE for $Erk$ when $Mek$ is fixed. (d) Comparison of SCM, SDE and ODE estimates of the ATE on $Erk$ when $Ras$ is fixed. } \label{fig:igfSCMvsSDE}
\end{figure}

\subsection{Case study 2: host response to viral infection}

\medskip \noindent {\textbf{Generating BEL causal model}}
The steps of the proposed \algref{query2bel} produced the qualitative causal model in~\figref{fig:covid}, and the corresponding BEL causal model $\mathbb{B}$, as follows.
In accordance with the inputs to  \algref{query2bel}, we defined the knowledge base $\mathbb{K}$ as the Covid-19 knowledge network automatically assembled from the \ac{cord19} document corpus using the \ac{INDRA} workflow.
We defined the cause $\mathbf{X}^{\mathrm{c}}$ as \ac{sil6r}, the effect $\mathbf{X}^{\mathrm{e}}$ as \ac{ards}, and the covariates $\mathbf{X}^{\mathrm{z}}$ as \ac{sars2} and \ac{toci}.
Therefore the causal query of interest was defined as $\mathbb{Q}=\left\{\aca{sil6r},\aca{ards}, \left\{\aca{sars2},\aca{toci}\right\}\right\}$. 

\algref{query2bel} line \ref{query_cause_to_effect} generated all pathways from \acl{sil6r} to \acl{ards}, resulting in
$kin(p(\aca{sil6r}))\rightarrow kin(p(\aca{il6stat3}))\rightarrow bp(\aca{il6amp})\rightarrow bp(\aca{ards})$, where $bp()$ is a biological process.
Next, line \ref{query_covariate_to_cause} generated all pathways from \acl{toci} to \acl{sil6r}:
$a(\aca{toci})\rightinhibits kin(p(\aca{sil6r}))$, where $a()$ is the dosage level of \acl{toci}.
We then generated all pathways from \acl{sars2} to \acl{sil6r}:
$ pop(\aca{sars2})\rightinhibits cat(\aca{ace2})\rightinhibits a(\aca{ang})\rightarrow kin(p(\aca{agtr1}))\rightarrow kin(p(\aca{adam17}))\rightarrow kin(p(\aca{sil6r}))$,
where $pop()$ is the viral load of \ac{sarscov2} and $cat()$ is the normal catalytic activity of \acl{ace2}.

Line \ref{query_covariate_to_effect} found no new branches from \acl{toci} to \acl{ards}.
Finally, we generated all pathways from \acl{sars2} to \acl{ards}, which resulted in three new branches
$pop(\aca{sars2})\rightarrow kin(p(\aca{prr})) \rightarrow kin(p(\aca{nfkb}))\rightarrow bp(\aca{il6amp}))$, 
$kin(p\aca{adam17}))\rightarrow p(\aca{egf}) \rightarrow kin(p(\aca{egfr}))\rightarrow kin(p(\aca{nfkb}))$, and
$kin(p(\aca{adam17}))\rightarrow kin(p(\aca{tnf})) \rightarrow kin(p(\aca{nfkb}))$.

\begin{figure}[t!]
\begin{center}
\includegraphics[width=0.48\textwidth]{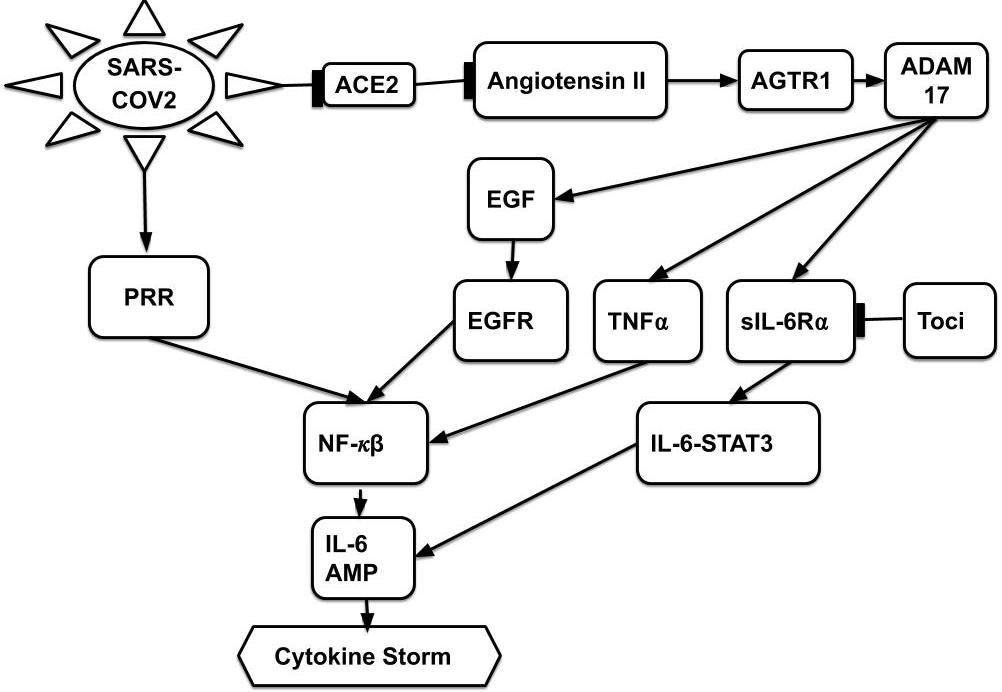}
\end{center}
\caption{\small  {\bf Case study 2: host response to viral infection}  
Pointed edges represent relationships of type {\it increase}; flat-headed edges represent relationships of type {\it decrease}. Nodes SARS-COV2 and \ac{toci} are external stimuli.}  \label{fig:covid}
\end{figure}

\medskip \noindent {\textbf{Observational data}} 
We simulated observational data from a ``ground-truth'' \acl{sigscm}, where the topology reflects the causal structure in~\figref{fig:covid}, and the parameters reflect our prior qualitative knowledge of the \ac{il6amp} pathway.
The root nodes \aca{sars2} and \acl{toci} were sampled from a Normal distribution with mean of $50$ and standard deviation of $10$.  
The non-root nodes were sampled from a sigmoid function as in \eqref{eqn:sigmoidFunc}.
Since we have prior qualitative knowledge that \ac{il6amp} is only activated due to simultaneous activation of \ac{nfkb} and \ac{il6stat3}, we set the threshold for activation above what could be achieved by \ac{nfkb} or \ac{il6stat3} alone.
Since we also know that \ac{toci} is a strong inhibitor of \ac{sil6r}, we set the inhibition coefficient to a large negative number.
The parameters of the sigmoid function were chosen to ensure that the variables were in the desired range of $0$--$100$.
Finally, we randomly generated two new individuals $\mathbb{D}^{new}$ with \acl{ards} $> 65$ to represent severely ill patients.
The first patient had a higher viral load of \ac{sarscov2} and received a lower dose of \ac{toci}.
The second patient had a lower viral load of and received a higher dose of \ac{toci}.

\medskip \noindent {\textbf{Estimation of individual-level treatment effect}}
\figref{fig:tocic} evaluates the SCM-based estimates of  the individual treatment effect of withholding treatment from two COVID-19 patients who were severely ill.
The distribution of the individual treatment effect obtained with the \ac{SCM} trained using \algref{beltoscm} was consistent with, but had a slightly larger variance then, the distribution of ITE obtained with the ``ground truth" \ac{SCM} with known weights.
Even though both patients had the same severity of illness prior to the intervention, patient B was estimated to have a more severe cytokine storm after \ac{toci} was withheld.

\figref{fig:tocicdirect} further compared the individual treatment effect obtained with the \ac{SCM} trained using \algref{beltoscm} with the average treatment effect estimated from the same model using forward simulation.
The distribution of the individual treatment effect was patient-specific and had smaller variance, thus illustrating the statistical efficiency of counterfactual inference.

\begin{figure}[t!]
\begin{center}
\begin{tabular}{ccc}
\includegraphics[width=0.23\textwidth]{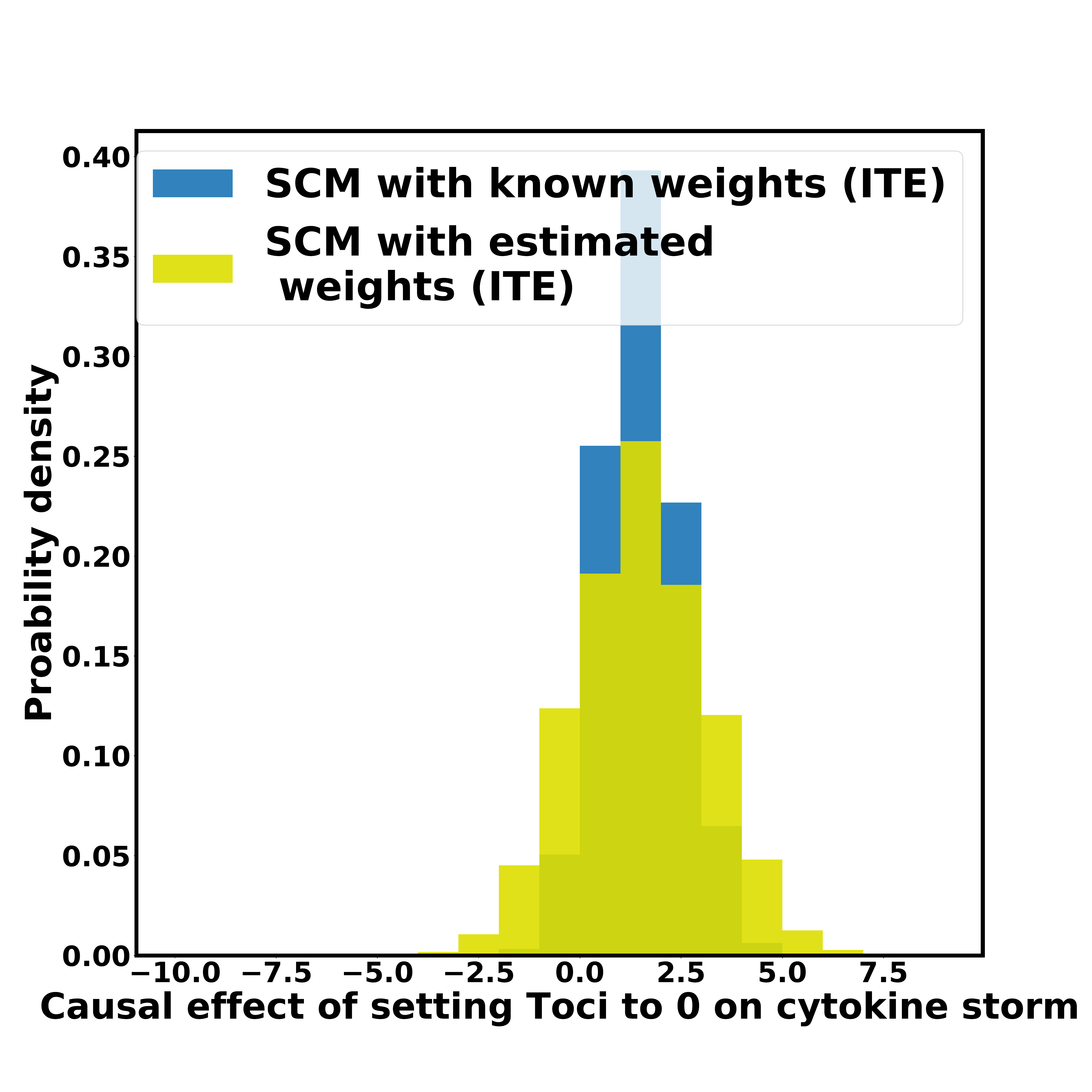}\label{fig:tocicounterfactual} &
\includegraphics[width=0.23\textwidth]{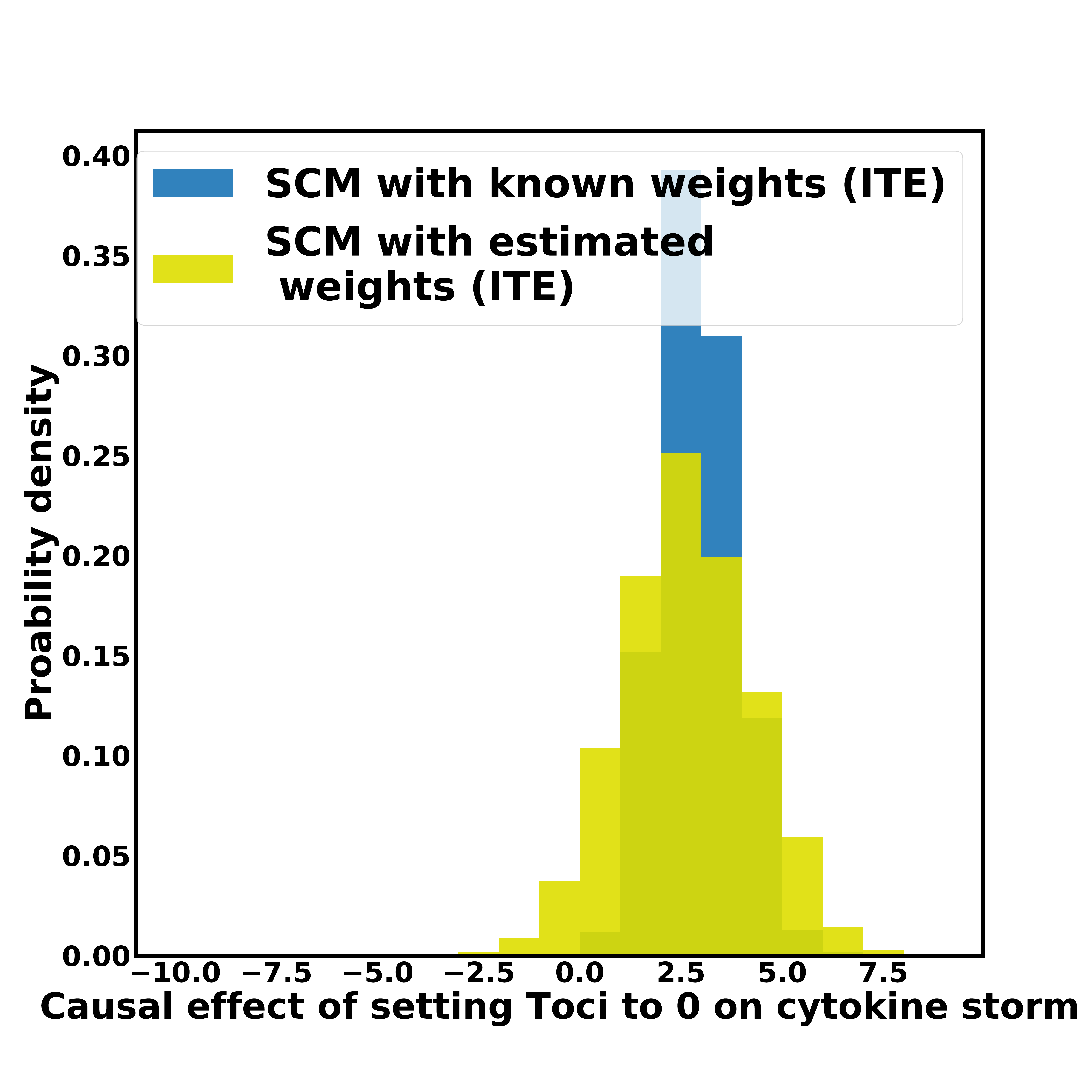}\label{fig:tocicounterfactual2}\\
(a) & (b)
\end{tabular}
\end{center}
\caption{\small \textbf{Case study 2: \ac{SCM}-based estimates of the \acf{ite} using \algref{ce}}.
  Blue histogram: the \ac{ite} estimated from the ground-truth \ac{SCM} using \algref{ce}.
  Yellow histogram: the \ac{ite} estimated from the \algref{beltoscm}-trained \ac{SCM} using \algref{ce}.
  (a) Patient has a high viral load and received a low dose of \acl{toci}.
  (b) Patient has a low viral load and received a high dose of \acl{toci}.
  Both patients were severely ill.
} 
    \label{fig:tocic}
\end{figure}

\begin{figure}[t!]
\begin{center}
\begin{tabular}{ccc}
\includegraphics[width=0.23\textwidth]{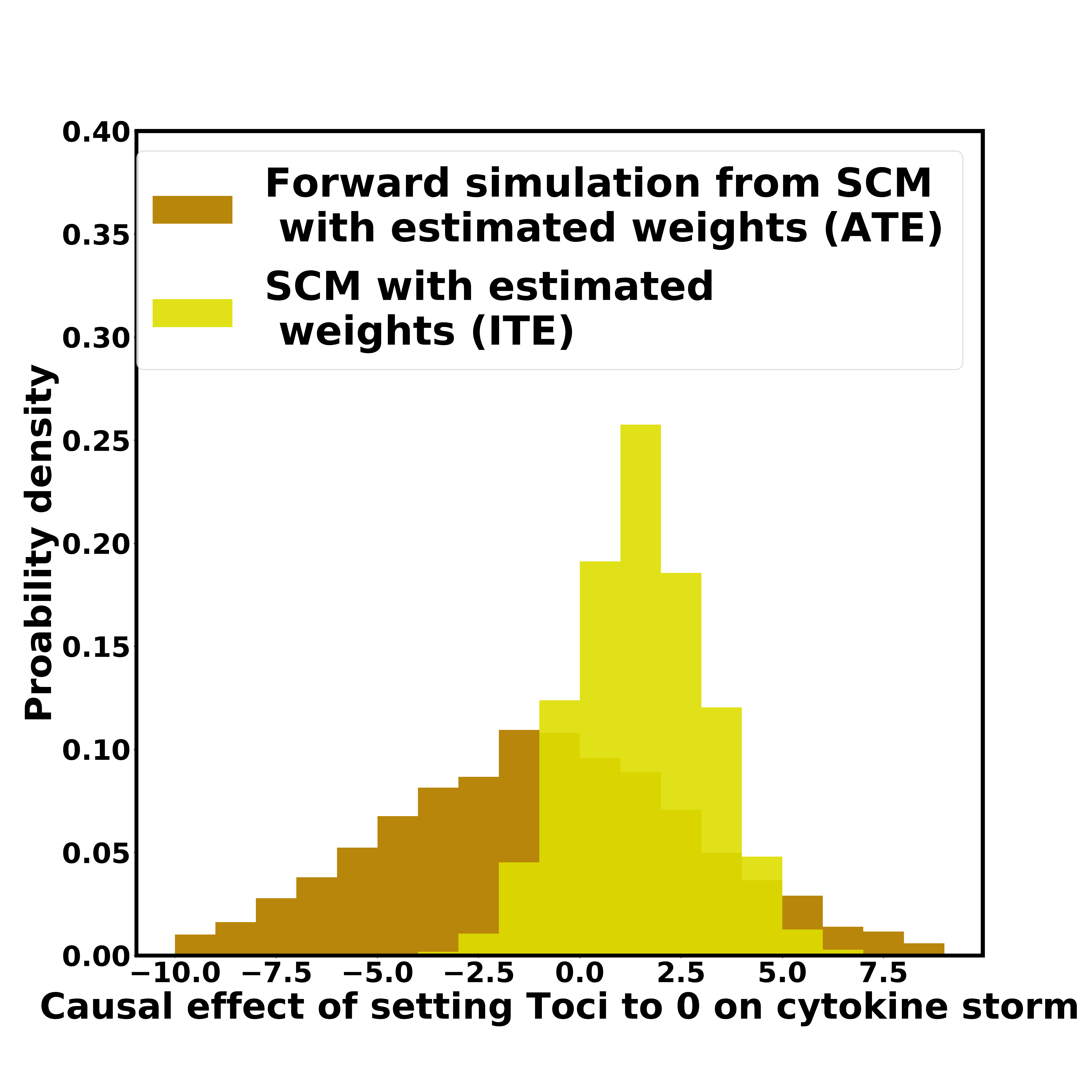} &
\includegraphics[width=0.23\textwidth]{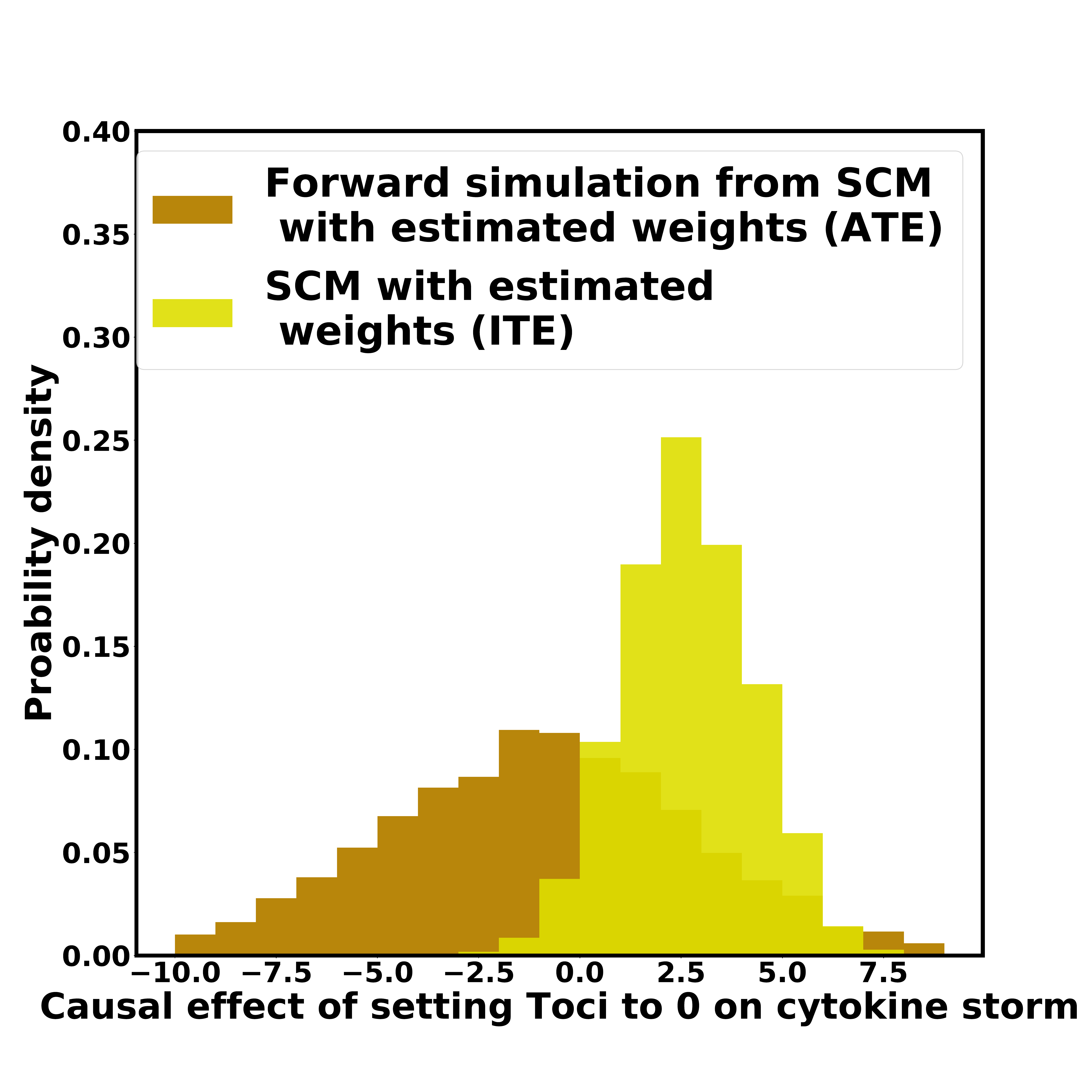}\\
(a) & (b)
\end{tabular}
\end{center}
\caption{\small \textbf{Case study 2: \ac{SCM}-based estimates of the \acf{ate} and of the \ac{ite} using \algref{ce}. }
  Yellow histogram: the ITE estimated using counterfactual inference.
  Brown histogram: the ATE estimated using forward simulation.
  (a) Patient has a high viral load and received a low dose of \acl{toci}.
  (b) Patient has a low viral load and a received a high dose of \acl{toci}.
  Both patients were severely ill.
}
\label{fig:tocicdirect}
\end{figure}
\section{Discussion}

We proposed a general approach that leverages structured qualitative prior knowledge, automatically generates a quantitative \ac{SCM}, and enables answers to counterfactual research questions.
In both case studies, the use of the \acl{BEL}  allowed us to leverage large repositories of structured biological knowledge to specify an \ac{SCM} and perform counterfactual inference in an automated manner, which would otherwise require a substantial manual effort.
The application to the IGF signaling system demonstrated the appropriateness of the underlying assumptions, and the accuracy of the results when compared to ODE- and SDE-based forward simulation.
The application to a study of host response to SARS-CoV-2 infection demonstrated the feasibility, versatility and usefulness of this approach as applied to an urgent public health issue.
In particular, the approach can help determine the amount of \acf{toci} required to reduce the severity of each individual's \ac{ards}.
Furthermore, in situations where treatment options are limited (as is the case SARS-CoV-2), counterfactual estimates enable a more precise conclusion regarding who would likely live without receiving the treatment, who would likely die even if they did receive the treatment, and who would likely live only after receiving the treatment.

The approach opens multiple directions for future research.
In particular, future work can extend the configurability of the \ac{bel2scm} algorithm by incorporating the rich type information in \ac{BEL}, mapping parent-child type signatures to functional forms such as post-nonlinear models, neural networks, mass action kinetics and Hill equations, and incorporating additional data types such as binary variables, categorical variables, and continuous variables with constraints on their domains.
In some cases, the variables in the model may not be directly observable, but may nonetheless be characterized by means of detectable molecular signatures.
For example, even if interferon signaling may not be directly observable using transcriptomics measurements, it may still be possible to infer the activity of interferon signaling by an upregulation of interferon stimulated genes (ISG).
Future work will focus on leveraging molecular signature databases to infer the activity of variables in the model, and on learning and/or evaluating the models using experimental data~\cite{Liu2019}.

We also note that experimentalists typically formulate biological processes as linear pathways (e.g., from $S_1$ to $Erk$ in the MAPK example) that can be effectively perturbed and measured in a laboratory setting.
Yet such boundaries of biological processes are quite arbitrary, and are therefore highly susceptible to confounders.
One way to address this issue is to search the knowledge graph for all common causes of variables in the causal model, use an identification algorithm~\cite{tikka_2017} to find the minimal valid adjustment set of the augmented model,
and then prune all common causes that do not contribute to that set.
This approach will require us to tackle the issues of parameter and causal identifiability in the presence of confounders.

In addition to unobserved confounders, the validity of causal inferences can be threatened by feedback loops, model misspecification, missing data, and out-of-sample distributions.
To address the possibility of feedback loops, we must consider the time scale at which these feedbacks reach steady-state:
fast timescale feedback loops can be addressed with the chain graph interpretation of SCMs~\cite{lauritzen_2002}\cite{sherman_2018};
intermediate timescale feedbacks can be addressed with non-recursive structural causal models~\cite{pearl2009causal};
slow timescale feedback loops can be handled by unrolling the structure of the SCM as is done with dynamic Bayesian networks~\cite{koller_2009}, or simply by representing the entire feedback loop as a biological process, as we did with \ac{il6amp}.
In the case of model misspecification, we will investigate the ability of counterfactual inference to improve the estimation~\cite{NIPS2019_9569}. 
For missing data, we can leverage causal inference recoverability algorithms that have been published recently\cite{nabi_2020}, and for handling out-of-sample distributions, we can leverage recent results applying causal inference to the problem of external validity~\cite{bareinboim_2016}.
Future work will focus on addressing these threats to validity when applied to real biological data.


\ifCLASSOPTIONcompsoc
  \section*{Acknowledgments}
\else
  \section*{Acknowledgment}
\fi

This work was supported by funds from the PNNL Mathematics and Artificial Reasoning Systems Laboratory Directed Research and Development Initiative. Knowledge curation environments were provided by BioDati.com and Causaly.com.  We would also like to acknowledge Jessica Stothers and Rose Glavin at CoronaWhy.org and Marek Ostaszewski at the COVID-19 Disease Map Initiative for providing valuable feedback about the IL6-AMP model.



\bibliographystyle{IEEEtran}

\bibliography{main}

\end{document}